\documentclass[12pt,preprint]{aastex}

\usepackage{graphicx}
\usepackage{epstopdf}
\usepackage{url}
\usepackage{amsmath}
\usepackage{rotating}
\usepackage{longtable}
\usepackage{multirow}
\usepackage[caption=false]{subfig}

\newcommand{\kepler}{\emph{Kepler}}

\slugcomment{Submitted to the Astrophysical Journal}

\shorttitle{Measuring Transit Signal Recovery}
\shortauthors{Christiansen et al.}

\begin{document}


\title{Measuring Transit Signal Recovery in the Kepler Pipeline I: Individual Events}

\author{Jessie L. Christiansen$^1$}
\author{Bruce D. Clarke$^1$}
\author{Christopher J. Burke$^1$}
\author{Jon M. Jenkins$^1$}
\author{Thomas S. Barclay$^2$}
\author{Eric B. Ford$^3$}
\author{Michael R. Haas$^1$}
\author{Shawn Seader$^1$}
\author{Jeffrey Claiborne Smith$^1$}
\author{Susan E. Thompson$^1$}
\author{Joseph D. Twicken$^1$}
\email{jessie.l.christiansen@nasa.gov}
\affil{$^1$SETI Institute/NASA Ames Research Center, M/S 244-30,  Moffett Field, CA 94035}
\affil{$^2$Bay Area Environmental Research Institute/NASA Ames Research Center, M/S 244-30, Moffett Field, CA 94035}
\affil{$^3$Astronomy Department, University of Florida, 211 Bryant Space Sciences Center, Gainesville, FL 32111, USA}




\begin{abstract}

The \kepler\ Mission was designed to measure the frequency of Earth-size planets in the habitable zone of Sun-like stars. A crucial component for recovering the underlying planet population from a sample of detected planets is understanding the completeness of that sample---what fraction of the planets that could have been discovered in a given data set were actually detected. Here we outline the information required to determine the sample completeness, and describe an experiment to address a specific aspect of that question, which is the issue of transit signal recovery. We investigate the extent to which the \kepler\ pipeline preserves individual transit signals by injecting simulated transits into the pixel-level data, processing the modified pixels through the pipeline, and comparing the measured transit signal-to-noise ratio (SNR) to that expected without perturbation by the pipeline. We inject simulated transit signals across the full focal plane for a set of observations of length 89 days. On average, we find that the SNR of the injected signal is recovered at $MS = 0.9973(\pm0.0012)\times BS - 0.0151(\pm0.0049)$, where $MS$ is the measured SNR and $BS$ is the baseline, or expected, SNR. The $1\sigma$ width of the distribution around this correlation is $\pm2.64\%$. We discuss the pipeline processes that cause the measured SNR to deviate significantly from the baseline SNR; these are primarily the handling of data adjacent to spacecraft re-pointings and the removal of harmonics prior to the measurement of the SNR. Finally we outline the further work required to characterise the completeness of the \kepler\ pipeline.

\end{abstract}


\keywords{techniques: photometric --- methods: data analysis --- missions: Kepler}

\section{Introduction}
\label{sec:intro}

The \kepler\ Mission is a NASA Discovery Program mission designed to examine the population of planetary systems using high precision photometric observations. The primary goal of \kepler\ is to measure $\eta_{\Earth}$, the frequency of Earth-size planets in the habitable zone of Sun-like stars. It was launched in 2009, and since then has been nearly continuously monitoring the brightness of $\sim$160,000 stars in 29.4-minute integrations, in a fixed field of view in the constellation Cygnus, looking for the periodic dimmings indicative of transiting planets \citep{Borucki2011a, Borucki2011b, Batalha2012}.

In order to determine the parent population of planets from the sample of detected planet candidate events, we are required to make a set of assumptions that must be carefully identified and justified in each analysis. Borucki et al. (2011b; referred to as B11 for the remainder of this paper), \citet{Youdin2011}, \citet{Howard2012}, \citet{Dong2012} and \citet{Fressin2013} describe initial analyses of the published \kepler\ planet candidate lists and preliminary attempts to constrain the underlying planet distribution. Two significant components of the analyses about which our knowledge is continuing to mature are the completeness of the planet sample (i.e., the false negative rate) and the reliability of the planet sample (i.e., the false positive rate). 

The false positive rate is a measure of the fraction of the planets in a given candidate list that are believed to be either astrophysical false positives, arising from eclipsing binaries (either star-star or star-planet) along the line of sight or bound in a system with a third body, or instrumental false positives, arising from spurious signals in the data which mimic a real transit signal. The false positive rate of the \kepler\ planet candidates published to date (see Burke et al. in prep, for the current list) has been investigated by several independent teams with some disagreement. \citet{Morton2011} assumed an intrinsic planet occurrence rate of 20\% for close-in planets and used Galactic stellar population models to determine an \emph{a priori} false positive rate of $\sim$10\% from the B11 candidate list. \citet{Santerne2012} performed high-resolution spectroscopy on a sample of bright, close-in giant planet candidate hosts, and found a false positive rate of $34.8\pm6.5$\%. This is even more discrepant with the \citet{Morton2011} result than first appears, since Figure 7 of \citet{Morton2011} indicates that the predicted false positive rate for Jupiter-sized candidates is close to 0\%. \citet{Colon2012} performed multi-colour transit photometry on a small number of close-in smaller planet candidates ($<5$ R$_{\Earth}$) and also found a false positive rate inconsistent with the \citet{Morton2011} result (2 of the 4 candidates they studied were determined to be false positives). \citet{Mann2012} studied a sample of bright, late-type \kepler\ targets, including many unclassified targets, and found that $96\pm1$\% were giant stars, which points to a 40\% misclassification rate in the Kepler Input Catalog (KIC; Brown et al. 2011), which was optimised for solar-like stars and included caveats for cool stars. This systematic increase in the radii of the planet candidate host stars would result in a corresponding increase in the planet candidate radii, pushing some of the candidates into the stellar size regime. However, since the reduced transit depth selects against detecting planets around giant stars, we do not expect the Kepler planet candidate hosts to be misclassified at the same rate, although we do expect an impact on the false positive rate. 

It is evident that the current false positive rate for the vast majority of the planet parameter space in the \kepler\ planet candidate lists is not well determined. We are currently working on a false positive analysis exploiting the exquisite precision with which we can measure the position of the flux from the target, which can change subtly during transit depending on the location of the target star and the location of the source of the transit signal (Bryson et al. in prep). 

The false negative rate of a given transit survey has typically been directly measured by the injection of simulated transit signals into the observed data and measurement of the subsequent rate of discovery. Studies by \citet{Burke2006}, \citet{Croll2007}, \citet{Weldrake2008}, \citet{Ballard2010} and \citet{Ballard2011} used extensive Monte Carlo simulations of transit injection and recovery to constrain the likelihood of a given transit signal being recovered in their analyses. Thus far there has been no direct measurement of the false negative rate in the \kepler\ planet candidates. \citet{Youdin2011} and \citet{Fressin2013} estimated the rate empirically for their analyses. Batalha et al. 2012 (referred to as B12 for the remainder of this paper) showed that the number and distribution of additional planet candidates detected in their sample compared to the previous catalogue was significantly higher and systematically different from what would be expected from a simple signal-to-noise extrapolation of the distribution of planet candidates listed in the previous B11 catalogue. This was partly due to some incompleteness in the B11 catalogue, where planets with a signal above the detection threshold in the original data were not detected, and partly due to the intervening improvement of the \kepler\ pipeline. We expect the B12 catalogue to also suffer from some degree of incompleteness for various reasons that will be discussed below. Indeed, several teams identified planet candidates that were not included in the B12 catalogue using the same or shorter observing baseline \citep{Fischer2012, Huang2012, Ofir2012}, though in many of these cases the \kepler\ pipeline identified these signatures, but they were not subsequently promoted to planet candidate status.

This work presents the first results of an experiment to directly measure the preservation of transit signals injected in the pixel-level data on a star-by-star basis in the \kepler\ pipeline (i.e. the extent to which transit signals are preserved over the course of the data reduction). This is an essential ingredient in calculations of the planet candidate completeness, since it has been typically been assumed to be 100\% in efforts to date. We investigate the validity of this assumption by injecting simulated transit signals into the \kepler\ pixels and processing the pixels through the pipeline in the same manner as the original data, and examining the measured signal-to-noise ratio (SNR) of the injected signals. In Section \ref{sec:pipeline}, we summarise the processes in the \kepler\ pipeline, and discuss those which might impact transit recoverability. In Section \ref{sec:design} we describe the experimental design and execution, in Section \ref{sec:results} we present the results, and in Section \ref{sec:discussion} we discuss the implications. Section \ref{sec:conclusion} summarises the conclusions and describes future work.


\section{The Kepler Science Pipeline}
\label{sec:pipeline}

The \kepler\ Mission design, performance and data products have been described in several papers \citep{Koch2010, Borucki2010}; we direct the reader there in order to familiarise themselves with the concepts and terminology used in this paper. To understand the method by which we inject simulated transits, and the processes that might perturb their recovery, it is important to first understand the data reduction pipeline. The pipeline has also been described in detail in a series of papers; for an overview see \citet{Jenkins2010a} and Figure 1 therein; we briefly review the processes here.

The Calibration (CAL) component, described by \citet{Quintana2010}, calibrates the raw pixels from each CCD channel\footnote{\kepler\ began with 84 operating channels; after $\sim$8 months of operation, one module (comprising four channels) failed, and we are currently operating with 80 channels.}, including pixels from the target stars, background pixels and collateral pixels. The latter are processed first, one set per channel, to determine the black (bias), dark and smear (since \kepler\ operates without a shutter) levels for that channel. All pixels on the channel are then calibrated using these levels. Corrections are also applied for the non-linearity, gain, electronic undershoot and flat-field variations, which are also channel-dependent. 

The Photometry Analysis (PA) component of the pipeline, described by \citet{Twicken2010a}, constructs the initial flux time series for each target star. First, argabrightenings \citep{Witteborn2011} and cosmic rays are identified and removed from the calibrated pixels. Then the calibrated, cleaned, background pixels are used to generate a two-dimensional background polynomial surface for each channel, and the background flux is calculated from the polynomial at the position of each pixel and subtracted from the flux value in that pixel. The flux time series (light curve) for each target is then generated by summing, for each cadence, the corrected flux value in each pixel in the optimal aperture defined for that target \citep{Bryson2010}.

The Pre-search Data Conditioning (PDC) component, initially described by \citet{Twicken2010b} and updated in \citet{Smith2012} and \citet{Stumpe2012}, fits and removes systematic signals from the light curves that are common to many stars on a given channel. See Figure 1 from \citet{Stumpe2012} for examples of the types of systematics that are corrected, some of which are described in Sections \ref{sec:distortion} and \ref{sec:masking} below. On a given channel, each light curve is correlated with every other light curve. For purely astrophysical signals, we have no expectation for correlations between light curves, which are much more likely to arise as a result of on-board systematics. We rank the light curves by their degree of correlation and select the first 50\%, which are the most highly correlated. We then use singular value decomposition (SVD) on the correlated light curves to create a set of cotrending basis vectors (CBVs). Using a Bayesian Maximum A Posteriori approach (MAP; see \citet{Smith2012} for details), we generate a correction for each light curve from the set of CBVs, using priors based on the locations and brightness of neighbouring targets.

In PDC, we also detect and attempt to correct Sudden Pixel Sensitivity Drop-outs (SPSDs). These are sudden decreases in the sensitivity of a given pixel, likely due to damage from a high-energy radiation event such as a cosmic ray. After the event, the sensitivity of the pixel is  permanently degraded \citep{Jenkins2010a}.

The Transiting Planet Search (TPS) component, described by \citet{Jenkins2002}, \citet{Jenkins2010b}, \citet{Tenenbaum2012a} and \citet{Tenenbaum2012b} searches the corrected light curves for periodic box-shaped signals. The data undergoes additional detrending before we apply a wavelet whitening filter, formed by decomposing the flux time series into a set of separate timescales using wavelets \citep{Jenkins2002}. We apply the same whitening filter to the box-shaped trial transit signal for which the whitened light curve will be searched. We can then generate a time series for each trial transit duration, the same length as the light curve, which contains for each cadence the detection statistic for that trial transit duration at that cadence. The detection statistic is the significance of the correlation of the whitened light curve with a whitened transit signal centered on that cadence, compared to the noise estimated for that cadence, i.e. it is the SNR of the cadence. We refer to these as the Single Events Statistics (SES) time series for the remainder of this work, and it is these detection statistics that we are investigating in this study. We then fold the SES time series at a set of trial periods, from a minimum of 0.5 days to a maximum of the duration of the light curve being searched, and generate a set of Multiple Event Statistics (MES) \citep{Jenkins2002}. In order for a given target to be considered a Threshold Crossing Event (TCE), the target must pass a set of tests, the most significant and straight-forward of which is that the maximum MES must be $\ge7.1\sigma$. See \citet{Tenenbaum2012a} and \citet{Tenenbaum2012b} for detailed description of the test criteria used to generate TCEs previously. Once a target passes the tests it is considered a TCE, and passed to the final component of the pipeline.

The Data Validation (DV) component, described by \citet{Wu2010}, performs a suite of diagnostic tests on the set of TCEs that are detected by TPS. These include tests on the light curve, such as examining the consistency of the depths of the putative transit signals and examining the fit of a physical transit model to the data. There are also tests on the spatial location of the centre of the flux in the target aperture, both in and out of transit, which is a powerful tool for diagnosing astrophysical false positives due to blended background eclipsing binaries. In this work, we are concerning ourselves only with the investigation of the detection statistics as reported by TPS; however, our long term plan is also to quantify the performance of the DV tests, especially in the area of identifying false positives.

\subsection{Signal Distortion}
\label{sec:distortion}

There are many processes by which a transit signal, present in the raw flux time series as measured by \kepler, could be distorted from its original shape and depth. One is aperture errors and losses due to changing amounts of flux falling into the target aperture; as described above, we integrate the flux for a given cadence over a fixed set of pixels. The \kepler\ spacecraft has very stable pointing (3 milli-arcseconds in 15 minutes, Koch et al. 2010), but jitter and drift in the pointing will cause a different fraction of the total flux from the target to fall within the fixed set of pixels in each cadence. A more significant effect is the change in focus that arises from the changing thermal environment of the \kepler\ spacecraft. As the focus changes, the pixel response function (PRF; Bryson et al. 2010) widens or narrows across the field of view (see the \kepler\ Data Characteristics Handbook, Christiansen et al. 2012 for examples), which again changes the fraction of the flux from the target falling within the fixed aperture. The focus change also affects the plate scale, such that the centre of the flux from a given target changes position with time. The focus change is dominated by a yearly seasonal variation as the spacecraft orbits the Sun while continuing to point at the same field of view. Perhaps more vexing are the focus changes due to thermal transients after spacecraft re-pointings. The historic pointing and focus changes for a given quarter have been shown graphically in the Data Release Notes\footnote{The Data Release Notes can be found at the MAST website: \url{http://archive.stsci.edu/kepler/data_release.html}} for that quarter. Another aperture error is caused by differential velocity aberration (DVA), where the positions of the targets within the field of view change relative to each other with time due to the relative motion of the spacecraft with respect to the field of view, up to 0.6 pixels over 90 days in the corners of the field of view. As the position changes, so does the fraction of the flux falling into the fixed aperture, producing errors similar to the pointing errors. 

\begin{figure}
\plotone{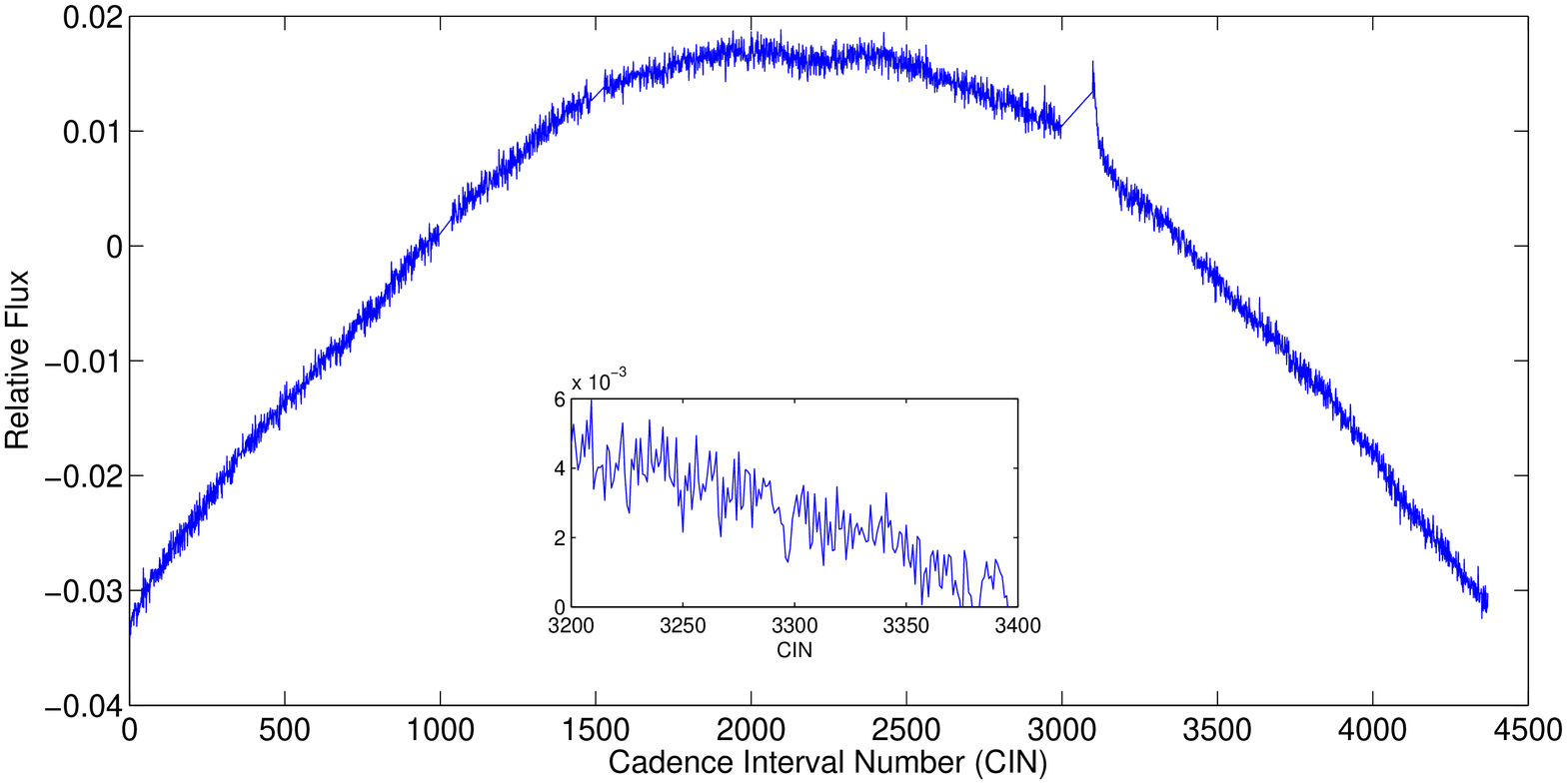}
\plotone{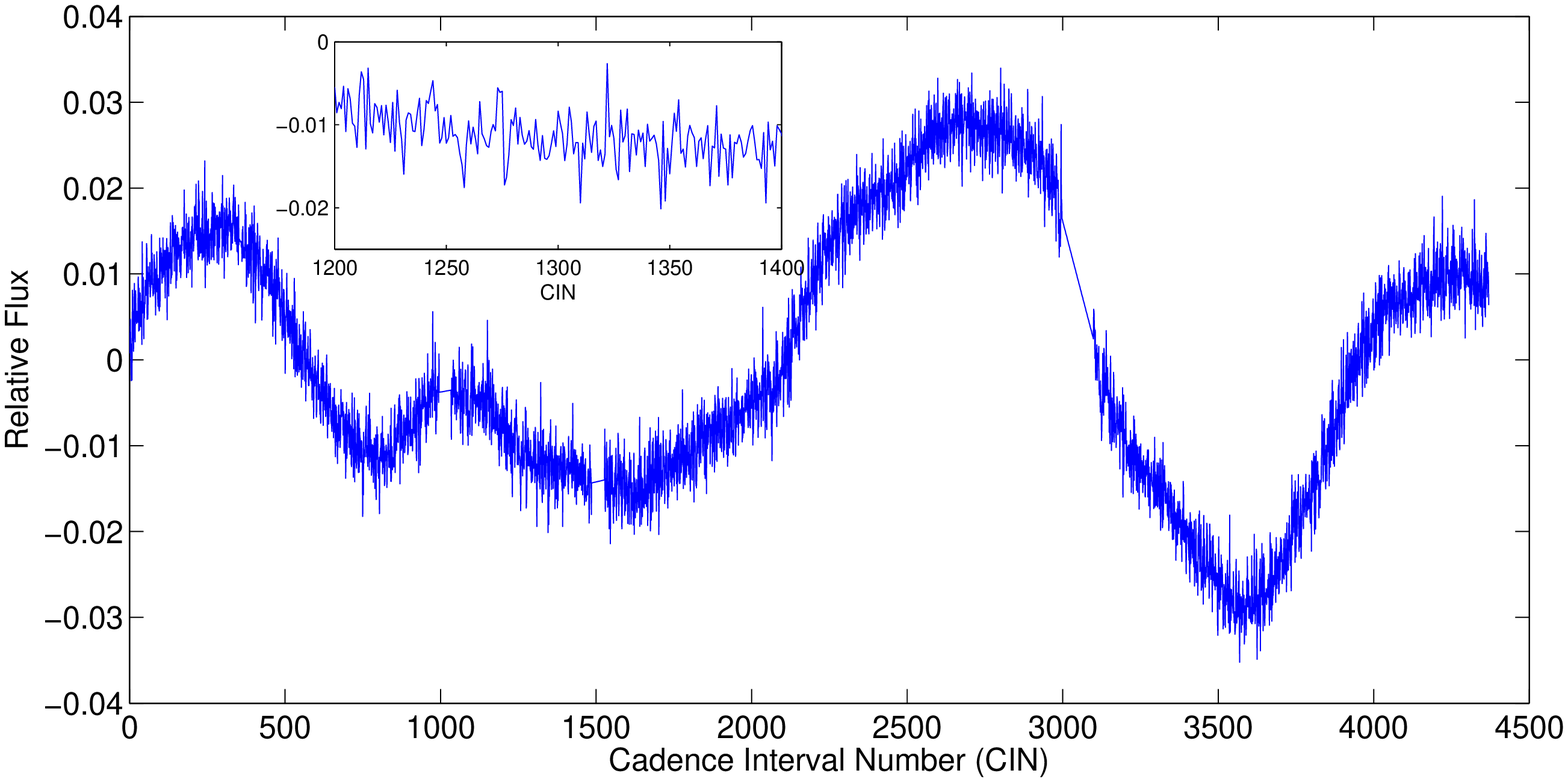}
\caption{Example CBVs from Quarter 3. \emph{Upper:} A strong CBV for module 14, output 2. The inset shows a transit-like feature near Cadence Interval Number (CIN) 3300. \emph{Lower:} A moderately strong CBV for module 14, output 2. The inset shows a series of negative outliers.}
\label{fig:cbv_example_1}
\end{figure}

A second potential source of distortion is the PDC component of the pipeline. As described above, in PDC we generate a set of cotrending basis vectors for each channel, from the 50\% most highly correlated light curves on that channel, and use linear combinations of these CBVs to remove the systematics from the target light curves. The CBVs can contain features that mimic transit signals, either from large astrophysical signals in the sample of highly correlated light curves (although these should be mitigated by the entropy cleaner in PDC, Smith et al. 2012), or more commonly from systematics that mimic transits. Since we are trying to remove the features we see in the CBVs, assumed to be caused by systematics, then any features in common between the target light curve and the CBVs will alter the fit and in turn be altered by the fit. Figure \ref{fig:cbv_example_1} shows examples of CBVs exhibiting transit-like features, both single events and quasi-periodic. In practice, the entire cotrending basis vector is being fit to the entire light curve; therefore scaling the CBV to match a local signal in the light curve will typically degrade the fit elsewhere, a less favourable solution. However, the presence of a signal in the CBV at the same time as a signal in the light curve will perturb the final fit, and could ultimately decrease the depth of the signal, however slightly. Conversely, if an otherwise well-fitting CBV has a positive deviation at the same time as a transit signal in the light curve, the signal would be distorted in the direction of increased depth. Therefore, since transit signals comprise a small fraction of the total time span of the light curve, any systematic signal distortion is expected to be small; quantifying this statement is one of the main goals of this work.

\begin{figure}[h!]
\plotone{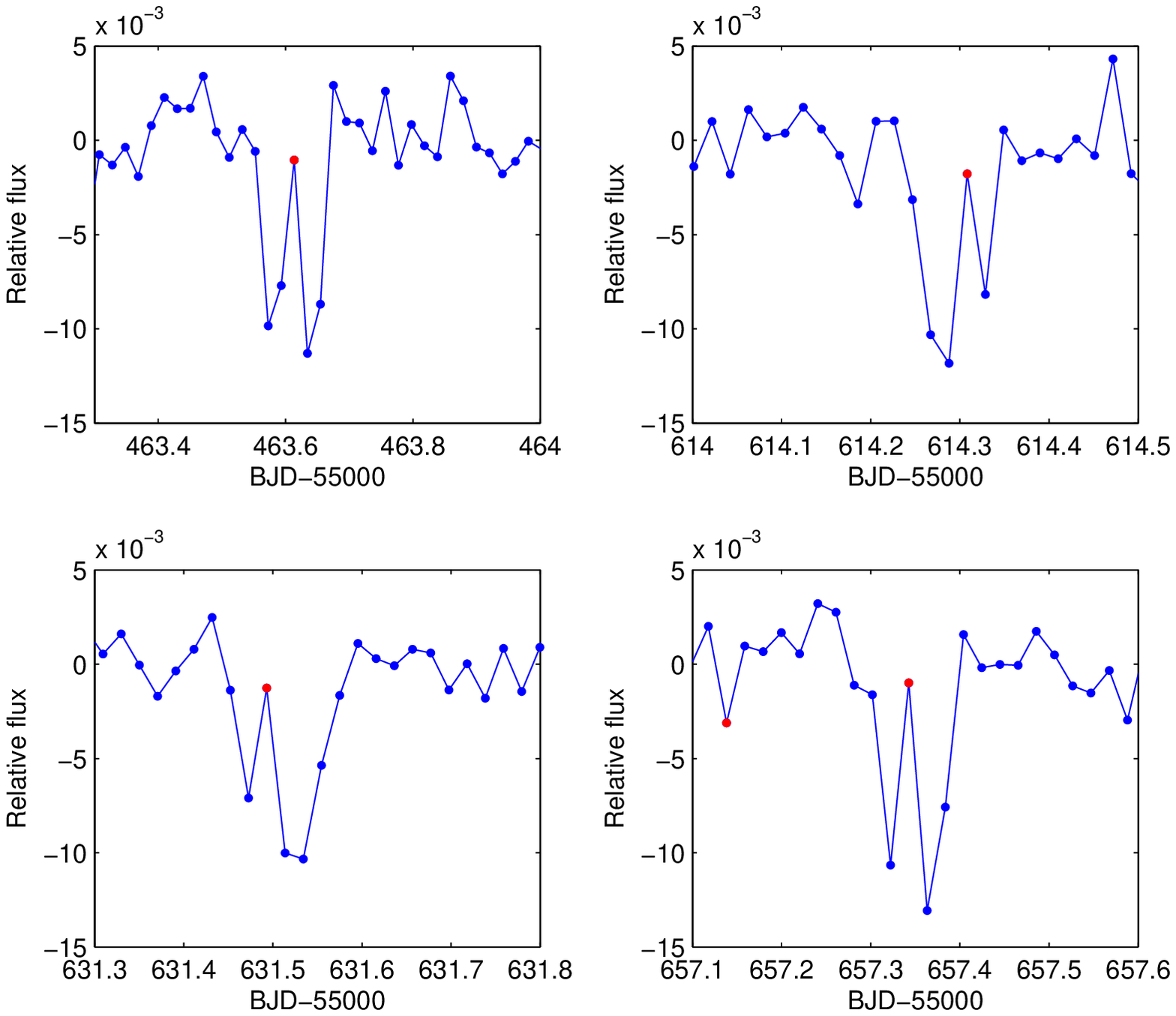}
\caption{Here we show four of the hundreds of transits of the candidate hot-Jupiter planet, KOI 667.01. The data shown are from the PDC-corrected light curve. The red data points indicate cadences which were mis-identified as cosmic rays by the pipeline and "corrected" to the cadence baseline.}
\label{fig:misidentified_CR}
\end{figure}

A third source of signal distortions are short duration deviations in individual flux time series, which are not common to multiple light curves and therefore are not amenable to being removed with the cotrending basis vectors, and are not corrected elsewhere in the pipeline. These include unidentified cosmic rays, and also the occasional cadence mis-identified as a cosmic ray and corrected back to the temporally local average flux. See Figure \ref{fig:misidentified_CR} for an example of the latter. When cadences occurring during transit events are displaced upwards in this fashion, it decreases the average depth of the signal, and ultimately decreases the measured signal strength of the event. Unidentified or improperly corrected SPSDs occurring during a transit will also alter the shape and detectability of the transit. By the very fact that these types of events are not identified, we cannot \emph{a priori} determine the fraction of transits that are likely to be affected. However by injecting simulated transit signals in the light curves, with their real distributions of cosmic rays and SPSDs, we can constrain the impact on the real signals by measuring the average detectability of the simulated signals.

\subsection{Signal Masking}
\label{sec:masking}

\begin{figure}[h!]
\plotone{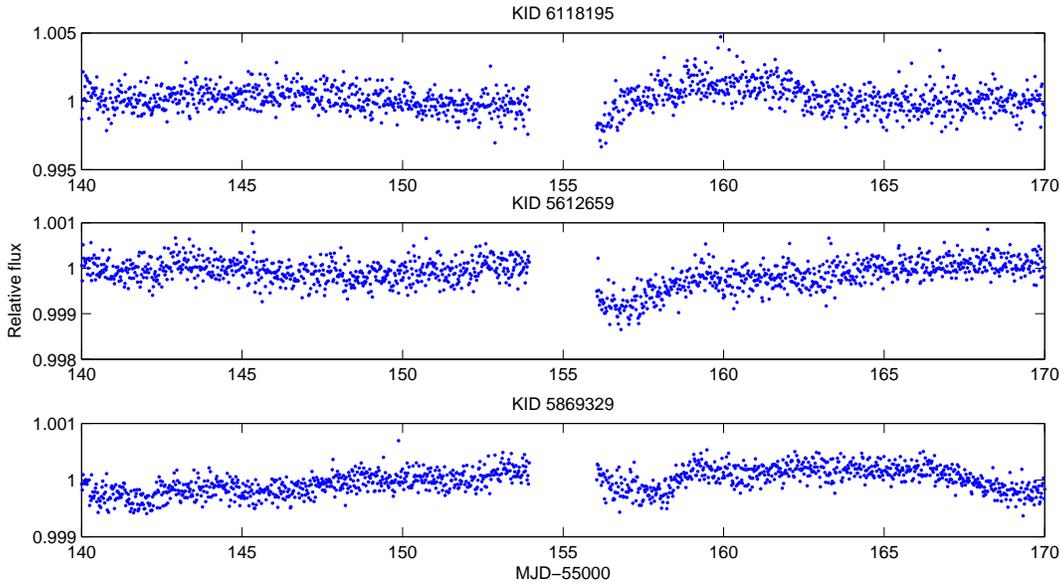}
\caption{This figure shows example corrected light curves from PDC that show thermal `hooks' after the data gap at day 155 between month 2 and month 3 of Quarter 3. The hooks are much more prominent in the uncorrected light curves, and the pipeline removes them almost entirely, but what is left behind can still mimic a valid transit detection.}
\label{fig:thermal_hook}
\end{figure}

\begin{figure}[h!]
\plotone{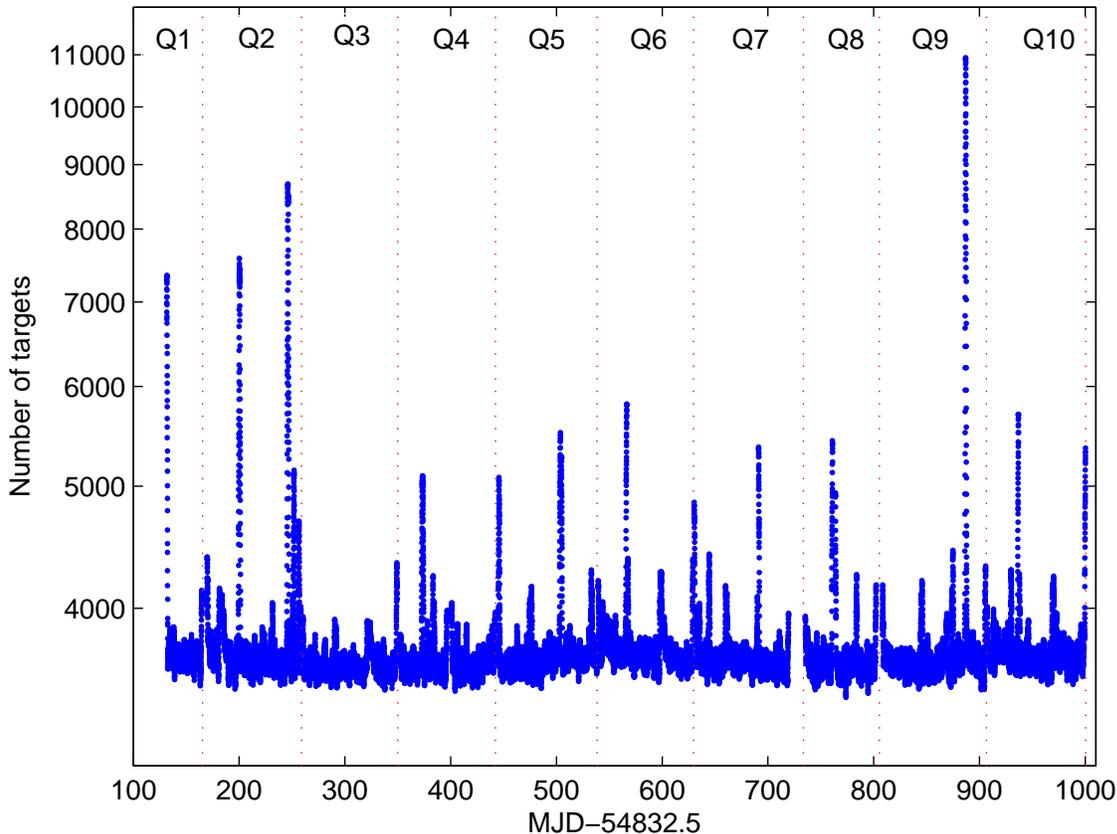}
\caption{Here we show the number of targets for each cadence in a Q1--Q10 pipeline run with the early version SOC 8.3 used for this study, for which that cadence contributed to the strongest signal detected for those targets. The largest numbers of targets occur on cadences that are affected by systematics mimicking transit events. The red dotted lines show the quarter boundaries, which are coincident with some of the peaks; the remainder of the peaks are largely caused by spacecraft re-pointings (during monthly data downlinks and safe modes). The final version of SOC 8.3 largely mitigated this problem by de-weighting cadences near the edges of long data gaps, as described in Section \ref{sec:suppression}.}
\label{fig:skyline_plot}
\end{figure}

Besides distorting the transit signals that are present in the light curves, uncorrected systematics can present as significant detections themselves, when compared to a box-shaped trial pulse. For instance, the thermal resettling after a spacecraft pointing change can often manifest itself in the light curves as slow increases in the measured flux over a day or so; see Figure \ref{fig:thermal_hook} for examples. One way to identify cadences that may be triggering detections due to uncorrected systematics is to count, for each cadence, the number of targets for which that cadence contributed to the strongest periodic signal. For uncorrelated astrophysical signals, we expect the events to be distributed randomly throughout the light curves, i.e. there should be no preferred cadences during which a significantly higher number of events occur. Figure \ref{fig:skyline_plot} shows an example distribution for the detections from a pipeline search of the Q1--Q10 data with the early version of SOC 8.3 used in this work. Most of the peaks in the distribution coincide with spacecraft re-pointings, which occur after data downlinks (monthly), safe modes and attitude tweaks. However, this only accounts for spurious detections arising as a result of systematics common to multiple targets---events like SPSDs, which are specific to a given target and randomly distributed in time, will not be evident in this figure.

This phenomenon, where the significance of the `detection' of a systematic could be higher than a real transit signal in the same light curve, is more severe in the first three catalogues released by the \kepler\ project. For these data sets, if the strongest signal found in a given target by TPS did not pass the additional tests, as is likely if it is caused by a systematic (see Tenenbaum et al. 2012a,b for further details), then TPS would not continue searching that target for weaker signals. Therefore, if there was a real transit signal in the flux time series that was weaker than the systematic, but stronger than the 7.1$\sigma$ threshold and therefore theoretically discoverable, it would  not have been detected by the pipeline. We can see evidence of this when examining the detection statistics of the planet candidates in B12---Fig. 7 of that paper shows that, instead of increasing towards 7.1$\sigma$ as expected, the distribution of signal strengths turns over at $\sim$10$\sigma$. This points to a population of systematics masking some fraction of real transit signals between 7.1 and 10$\sigma$.

In the fourth sample of planet candidates (Burke et al., in prep) and future releases, this problem is largely mitigated by a re-design of the pipeline. Now, instead of discontinuing searching if the strongest signal in a given flux time series does not pass the tests in TPS, we mask out the cadences that contributed to the spurious signal and perform a new search. We iterate this process until the strongest signal returned by the search passes all the tests or is under the 7.1$\sigma$ threshold, which should greatly improve the rate of recovery of real transit signals above the threshold. We caution that for completeness analyses, it will be important to take into account the reduced number of cadences that ultimately contribute to the detection. For a given orbital period of interest, we also expect that signal masking will decrease with the increasing baseline of cadences, since real signals will increase in significance with more events, as compared to spurious signals caused by systematics, which are not likely to continue lining up in the same way. In this work, we do not attempt to constrain the rate of signal masking in the released catalogues, but we include this discussion here for clarity and to guide future work.

\section{Experiment Design}
\label{sec:design}

We would like to assess the recoverability of a given transit signal in a given target light curve. Ideally, we would measure this by injecting the target flux time series with a grid of simulated transit signals, over a set of planet parameters of interest (e.g. size, orbital period), and for each signal, process the light curve through the \kepler\ pipeline and directly measure the detection statistics. However, due to the number of \kepler\ targets ($\sim$190,000 with at least some observations), and the number of observations per target (40,000 and growing), this is computationally infeasible, and we need to find a different solution.

Thus, we will assess the recoverability of a given transit signal in an \emph{average} target light curve, and thus reduce the question to ensemble statistics. This decreases the number of tests required by several orders of magnitude to the point that it becomes tractable. To achieve this, instead of running multiple simulations on each target, we inject each target light curve with a different transit signal arising from a randomly generated single planet candidate, described below, and then measure the recovered detection statistics on average.

Another way to reduce the computational burden is to reduce the observation baseline, since the number of searches performed by the pipeline increases as $N^2$, for $N$ observations. Therefore we perform this initial test using a single quarter of data---Quarter 3, which spanned 89 days from 2009 September 18 to 2009 December 16. Since we are largely concerned with the average signal distortion introduced by the processes in the pipeline, we can treat each separate transit event in the light curve as an independent statistical test of the distortion. This allows us to place the transit epochs unphysically close together in the flux time series and therefore to sample more of the systematics in a given flux time series, as long as the transit events are separated well enough in time so as to not mutually influence each other's measured SNR. The variance window over which TPS calculates the noise, in order to determine the significance of a detection, is 30 times the duration of the box pulse being tested; outside of this window, the local noise properties should not be influenced by neighbouring transits. For this investigation, we separate the injected transit epochs by 50 times their duration. In Sections \ref{sec:results} and \ref{sec:discussion}, we discuss pipeline processes that might be influenced by this relatively close separation of transit epochs.


\subsection{Transit model construction}
\label{sec:model}

We generate our injected transits using the \citet{Mandel02} model. The parameters of the model are constructed from three observable parameters: (1) The signal-to-noise ratio (SNR), which is randomly drawn from a uniform distribution between 2$\sigma$ and 20$\sigma$. This range allows us to examine any correlation between the distortion and initial baseline SNR; (2) The signal duration, which is randomly drawn from a uniform distribution between 1 and 16 hours. In the pipeline, we search for transit pulses with durations from 1.5--15 hours; (3) The phase of the first injected transit, which is randomly drawn from a uniform distribution between 0--1. In a following step, this is used with the separation of the transit events to calculate the initial epoch.

From a few starting assumptions and knowledge of the target star, we can use these observable parameters to determine the planet parameters required to generate the model. For this test, we assume circular orbits (eccentricity of zero) and central-crossing transits (impact parameter of zero). Eccentricity has only a very slight impact on the transit shape and therefore recoverability, and assuming circular orbits allows us to easily calculate the orbital period of the injected planet from the selected signal duration. On the other hand, the impact parameter has a significant effect on the shape and depth of the transit. However, we are primarily concerning ourselves with the question of distortion in the measured detection statistics: for an initial signal with 2--20$\sigma$ significance, what is the typical final detection statistic measured by the pipeline? Since the transit depth is a function of both the injected planet radius and the impact parameter, we need only allow one of these to be a free parameter to test the correlation. In addition, the processes described in Section \ref{sec:distortion} should not affect differently-shaped transits in systematically different ways. The possible exception is the potentially higher rate of cosmic ray misidentification in more v-shaped transits, although as shown in Figure~\ref{fig:misidentified_CR} this occurs in flat-bottomed transits as well. A recent update to the cosmic-ray identification algorithm is expected to improve this behaviour when the data are re-processed.

To generate the model, we use the stellar parameters (surface gravity, effective temperature, stellar radius and metallicity) from the KIC, and default to solar values for unclassified stars. The pipeline generates a noise estimate for use in measuring the SNR of a putative transit signal, called the Combined Differential Photometric Precision \citep{Christiansen2012}. For a given cadence and box-pulse duration, it is the depth of the box at that cadence which would produce an SNR of 1. We use the rms CDPP  previously calculated by the pipeline for the target flux time series, prior to the injection of simulated planets, for the duration which is closest to our model signal duration. We calculate the model transit depth, $\delta$, in $\sigma$ from the product of the rms CDPP and the model signal strength. From the transit depth, we can calculate the planetary radius, $R_p$, from $\delta = (R_p/R_{\star})^2$, where $R_{\star}$ is the stellar radius.

We estimate the orbital period from Eq. (1) of \citet{Gilliland2000}. We then calculate the semi-major axis, $a$, from Newton's modification of Kepler's third law, and the geometric ratio $a/R_{\star}$. We can then use the \citet{Mandel02} analytic transit model formalism to generate a physical transit model, centred at the calculated starting epoch, and repeated along the light curve at our pre-defined transit separation, which is 50 times the model transit duration, as described earlier. The model is generated at a sampling rate 30 times higher than the flux time series, and then re-sampled onto the final time stamps. We use the KIC magnitude for each target star to convert relative depth as calculated by the model into the total number of photoelectrons that need to be subtracted from the light curve. We subtract absolute numbers of photoelectrons instead of a relative numbers to account for the fact that the pixels comprising a given light curve may contain flux contributions from multiple targets. The simulated transit model is then used twice: firstly, it is injected into the calibrated pixels, as described below, which are then processed through the pipeline; and secondly, it is injected into the original `clean' flux time series at the end of the pipeline to generate a set of baseline, or expected detection statistics for a signal unmodified by the pipeline processing.


\subsection{Pixel-level transit injection}
\label{sec:pixels}

We inject the generated transit model into the calibrated pixels for the target in question. In order to determine the correct number of photoelectrons to be subtracted from each pixel, we must first measure the position of the target. We use a set of $\sim200$ bright, unsaturated, target stars on each channel to derive the conversion from celestial coordinates---Right Ascension (RA) and Declination (Dec)---to pixel coordinates for each cadence, fitting a polynomial to the KIC RA and Dec of the bright stars and their measured pixel location in that cadence. By injecting transit signals into every target, and increasing the number of events injected per target above expectations, we risk introducing noise into these derived polynomials that would not be present in a normal pipeline run. Therefore we use the ``motion'' polynomials derived from an identical `clean' pipeline run, without the transits injected. 

\begin{figure}[h!]
\plotone{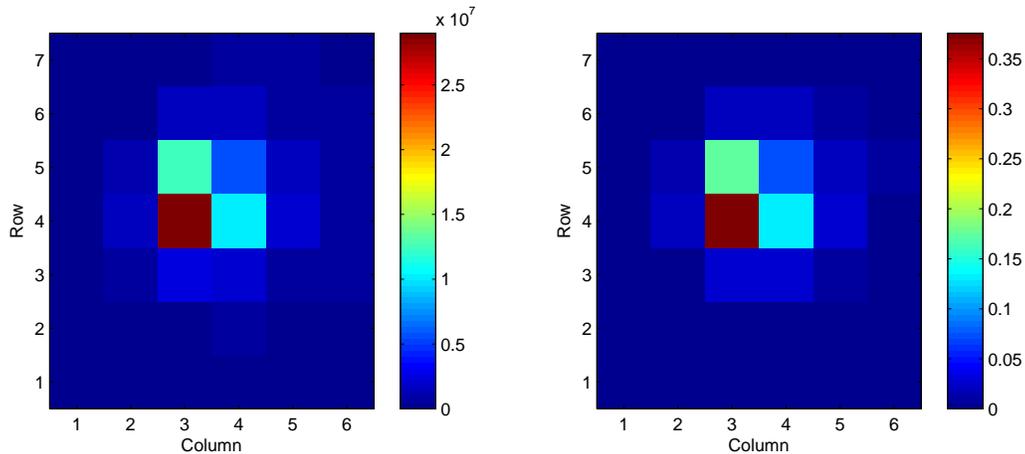}
\caption{\emph{Left:} The pixels surrounding isolated target KID~7757236, centred at local column 3, row 4. The flux units are in electrons/cadence. \emph{Right:} The model PRF interpolated onto the same position, in relative flux.}
\label{fig:PRF_isolated}
\end{figure}

\begin{figure}[h!]
\plotone{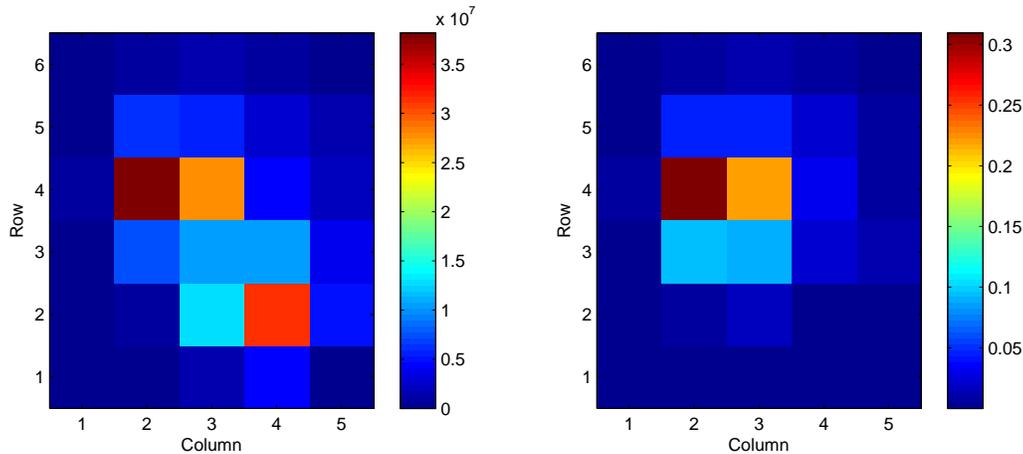}
\caption{\emph{Left:} The pixels surrounding target KID~7756621, centred at local column 2, row 4. Another slightly brighter source is centred at local column 4, row 2. The flux units are in electrons/cadence. \emph{Right:} The model PRF interpolated onto the same position, in relative flux, showing that some of the pixels for KID~7756621 contain contributions from the nearby source.}
\label{fig:PRF_crowded}
\end{figure}

Following the standard pipeline procedure, we use these polynomials to calculate the precise location of each target for each cadence using the KIC RA and Dec. The pixel response function (PRF) is a function of position, so for each target we generate a local PRF by interpolating between five PRF models to the derived location \citep{Bryson2010}. Figure~\ref{fig:PRF_isolated} shows an example scene for an isolated target, KID~7757236, with a \kepler\ magnitude $Kp=13.8$; the left panel shows the calibrated pixel values for a given cadence, and the right panel shows the interpolated PRF. We then render the modelled PRF onto the pixels that comprise the target star, which tells us the fractional contribution of each pixel to the total flux from that star. For each pixel we then subtract the appropriate fraction of the total number of photoelectrons required to model the transit at that time. We subtract from the flux instead of scaling the flux because individual pixels can have flux contributions from multiple targets, and we must only reduce the flux from the target of interest. This is particularly important for the false positive tests based on the change in the location of the centre of the flux in and out of transit---scaling the flux would not preserve the spatial information in the distribution of flux in the local scene. An example of this is shown in Figure \ref{fig:PRF_crowded}, which shows the local scene for target KID~7756621, with $Kp=13.4$, in the left panel. From the interpolated PRF for this target, shown in the right panel, we can see that several pixels are shared between the target (centred at local column 2, row 4), and a nearby source (centred at local column 4, row 2). 

\subsection{Pipeline processing}
\label{sec:pipelineorder}

These modified pixels are then processed through the pipeline as normal. The only departure from standard operations is that, like the motion polynomials, the cotrending basis vectors used in the correction of systematic errors are generated from a `clean' pipeline run. Again, this is to avoid corruption of the CBVs from the presence of many transits in every light curve, since the CBVs are generated from the data themselves. Of course, in reality there are some number of real transit events in a fraction of the light curves, however they should be uniformly scattered throughout and significantly outnumbered by light curves without transit events. In general, CBVs should not be affected by the presence of transits in a small fraction of the light curves, although as described before, those transits can be affected by the CBVs themselves during correction.

In summary, the final order of processing is that we run the calibrated pixels (the output of CAL) of Quarter~3 through PA, PDC, and TPS, without any modification, to generate the motion polynomials, the cotrending basis vectors, and the rms CDPP for each target. We then inject the simulated transit signals into the calibrated pixels, one planet for every target in the full focal plane (84 channels), and re-run the modified pixels through PA, PDC and TPS, utilising the previously generated information as described.

\section{Results}
\label{sec:results}

Of the 84 channels in the field of view, 80 channels completed the processing described in Section \ref{sec:pipelineorder}. Four channels (channels 23, 45, 62 and 67) failed for reasons unrelated to the transit injection and are not included in the subsequent analysis.

Our goal is to quantify the impact of the pipeline processing on the original injected signal. To achieve this, we inject the simulated transit signals in the pixel-level data, and compare the measured SNR (MS) to the expected SNR for that model. Note that when we inject the model into the real data, local noise will result in individual transit events having a range of measured detection statistics, independent of distortions caused by the pipeline processes, so we cannot simply compared the measured SNR to the simulated model SNR. To isolate the impact on the transit signal due to the pipeline from local correlated noise, we instead take the unmodified flux time series from the first pass through the pipeline, and immediately prior to the measurement of the SNR, inject the identical set of models to those that were processed all the way through the pipeline. This then generates the baseline SNR (BS), which represent how the signal would have been measured in the flux time series without any perturbation by the pipeline.

The impact of the pipeline processes is determined by comparing the measured SNR to the baseline SNR. In the comparison, we only consider the cadences that were modified by the injection of transits. We are also only considering targets for which the pipeline did not react strongly to the presence of many additional transits in the flux time series. This seems counter-intuitive, since we are investigating how the pipeline impacts transits; however in this study we are concerned with individual transit events, not the collective transit signal train. The motivation for the unphysical spacing of transit epochs is to increase the number of statistical tests, not to affect the behaviour of the pipeline. There are two ways the pipeline can react to the presence of many transits for which we filter here: the first is the SPSD detection algorithm, and we therefore exclude targets where the SPSD detections differed between the two runs. The second effect is on the target variability measured by PDC, which treats targets differently based on this measurement. We exclude targets where the measured target variability changed by $>$0.2. This typically led to the exclusion of several hundred targets per channel (20--30\% of the total number of targets); the final numbers for each channel are given in Table \ref{tab:all80channels}. Our long term goal is to examine recoverability of transit signal trains, as compared to single transits, at which point we will separate the transit epochs with realistic spacing and investigate the impact of these effects more thoroughly.

For the remaining targets, we want to identify the distribution of the measured SNR for a given baseline SNR. We first examine a set of individual channels in detail, in order to characterise the types of impact the pipeline could have on a real transit signal in the data. We then examine the distribution across the full set of channels.

\subsection{Individual channel results}

\subsubsection{Channel 1}
\label{sec:channel1}

\begin{figure*}[h!]
\centering
\includegraphics[width=0.45\textwidth]{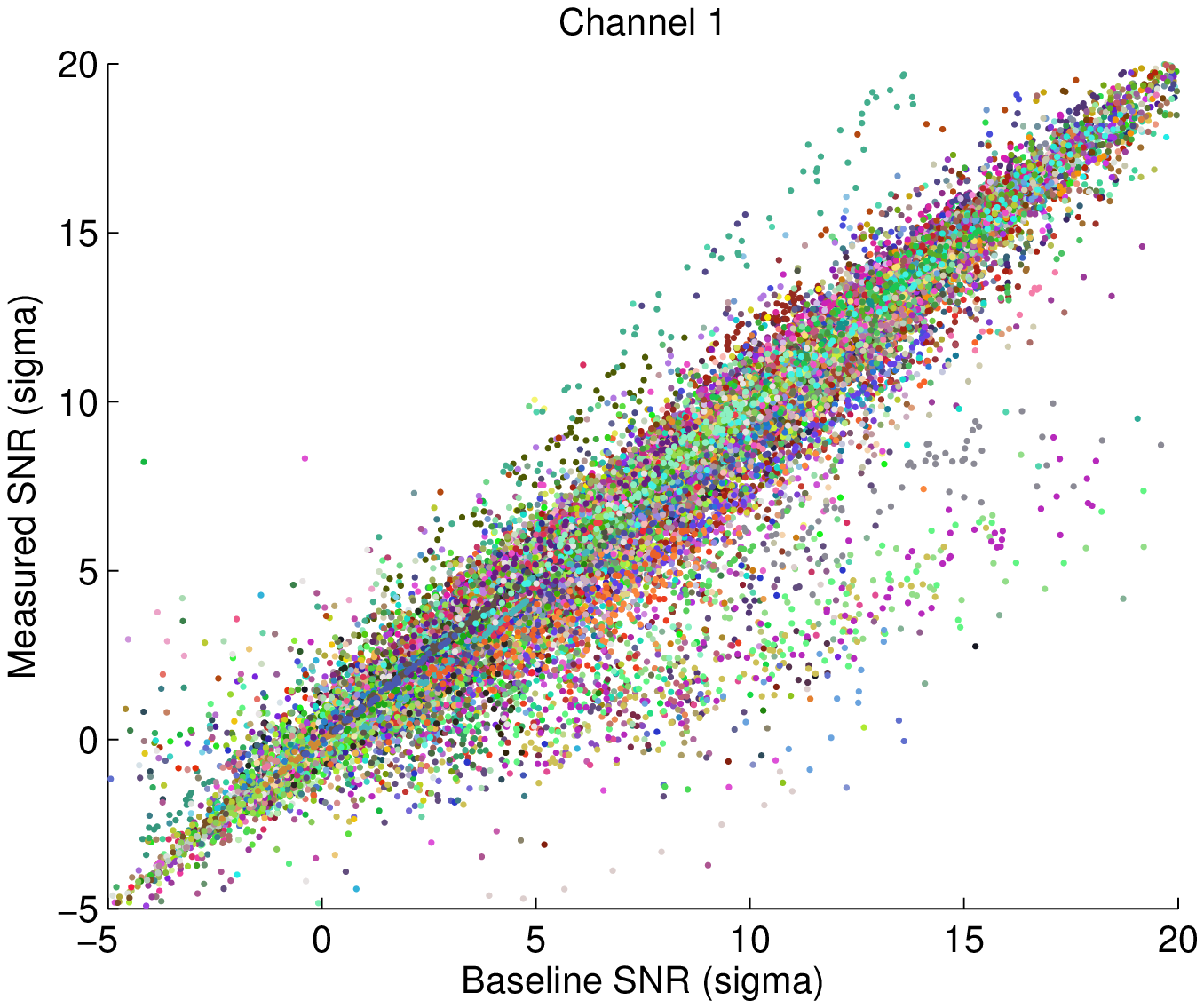}
\includegraphics[width=0.45\textwidth]{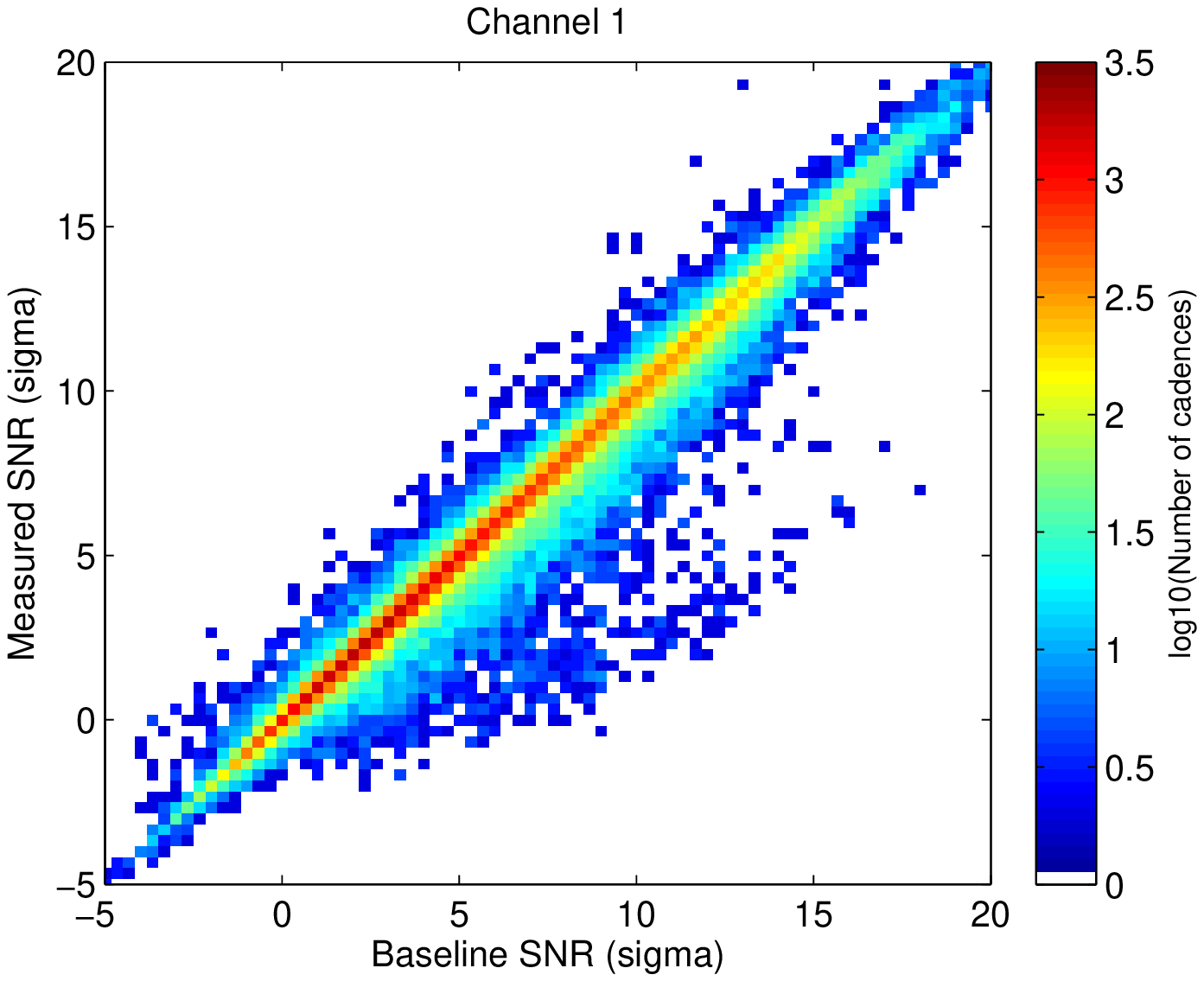}
\caption{The measured and baseline SNR for the 1130 targets on channel 1. \emph{Left panel:} Each cadence that was modified by the injection of a transit signal is plotted; all points associated with the same target are plotted in the same colour. \emph{Right panel:} A density plot the same data as the left panel, showing a very tight distribution around the 1:1 correlation (note the log scale in density).}
\label{fig:channel1density}
\end{figure*}

Figure \ref{fig:channel1density} shows the measured SNR plotted against the baseline SNR for the 1130 targets analysed on channel 1. The left panel shows each cadence plotted as a separate point. All points for a given target are the same colour (although the same colour can be used for multiple targets). The width of the scatter around the 1:1 correlation in this visualisation is misleading due to the number of cadences being plotted---the right panel shows the same data as a density plot, with a log scale in density, showing that the large majority of cadences are measured with an SNR extremely close to the baseline SNR. Performing a robust linear fit of the form $MS = a \times BS + b$, where $MS$ is the measured SNR of the transit single processed though the pipeline and $BS$ is the baseline SNR, both in units of sigma, to the data in the left panel yields coefficients of $a = 0.9970\pm0.0003$ and $b = -0.0218\pm0.0015$, indicating extremely high fidelity on average for the measured SNR to the baseline SNR. The fit is an iteratively re-weighted least squares fit using a bi-square weighting function.

\begin{figure*}[h!]
\centering
\includegraphics[width=\textwidth]{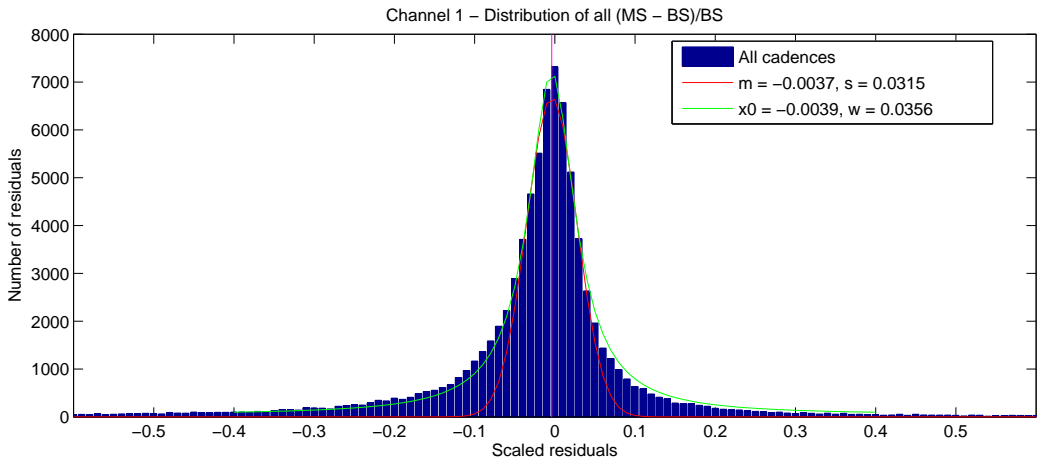}
\caption{The distribution of scaled residuals $(MS-BS)/BS$ for the data shown in Figure \ref{fig:channel1density}.}
\label{fig:channel1allresids}
\end{figure*}

Characterising the width and shape of the distribution of the residuals around this correlation allows us to determine the likelihood of recovering a measured SNR for a given baseline SNR. Figure \ref{fig:channel1allresids} shows a histogram of the fractional change in the measured SNR (($MS-BS)/BS$) from the baseline SNR for each cadence modified by transit injection. The core of the distribution ($|(MS-BS)/BS| < 0.06$) is well fit by a Gaussian form (shown in red), with a half-width half maximum (HWHM) of 0.0315. There are significant wings to the distribution however and as such we also show the fit of a Lorentzian form out to $|(MS-BS)/BS| < 0.4$ (shown in green), with a slightly wider HWHM of 0.0356. Therefore on average the measured SNR of the injected transit is recovered to within 3--4\%.

\begin{figure*}[h!]
\centering
\includegraphics[width=\textwidth]{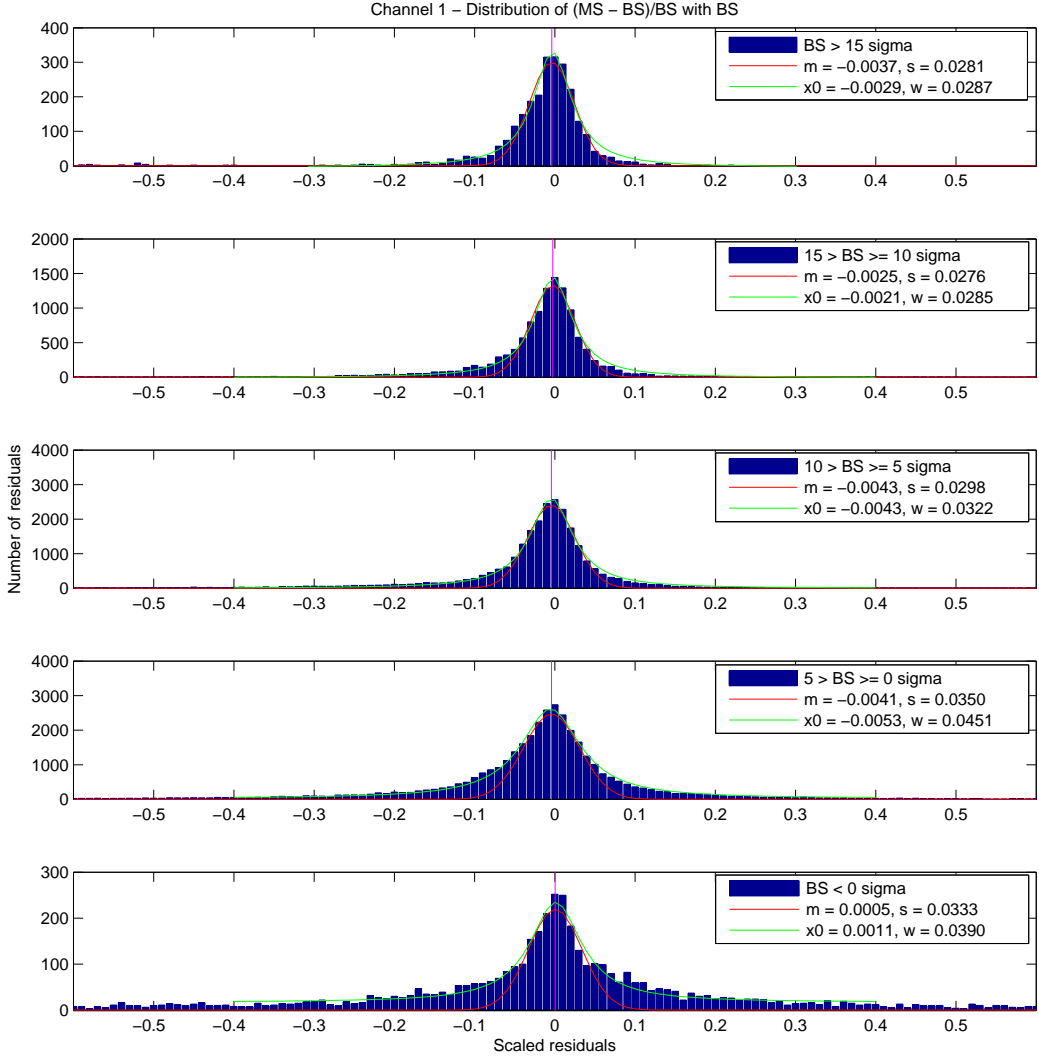}
\caption{The distribution of scaled residuals $(MS-BS)/BS$ from Figure \ref{fig:channel1allresids} in bins of decreasing baseline SNR from the top panel ($BS>15\sigma$) to the bottom panel ($BS < 0\sigma$) in bins of $5\sigma$.}
\label{fig:channel1resids}
\end{figure*}

To examine the dependence, if any, of the shape of the distribution on the baseline SNR, Figure \ref{fig:channel1resids} shows the same histogram of scaled residuals as Figure \ref{fig:channel1allresids} as a function of the baseline SNR. The five panels show data from the highest baseline SNR values ($>15\sigma$) to the lowest ($<0\sigma$), in bins of $5\sigma$. We again show the Gaussian fit to the core of the distribution ($|(MS-BS)/BS| < 0.06$) and the Lorentzian fit to the core+wings ($|(MS-BS)/BS| < 0.4$). The higher baseline SNR bins are well fit by the Gaussian form, however towards lower baseline SNR values the distribution widens, and we see that the aforementioned wings in Figure \ref{fig:channel1allresids} are dominated by the contribution from lower SNR values. At high values of the baseline SNR ($>10\sigma$), the half-width half maxima (HWHM) of the Gaussian and Lorentzian functions are similar (0.0281 and 0.0287 respectively for $BS>15\sigma$ and 0.0276 and 0.0285 respectively for $15\sigma>BS>10\sigma$); at low values the HWHM of the Lorentzian function is slightly higher (0.0390 for the lowest baseline SNR bin, compared to 0.0333 for the Gaussian fit). While the distribution widens from $\sim3\%$ at higher values to $\sim4\%$ at lower values, we see no significant change in the centre of the distribution from high baseline SNR to low baseline SNR.

Besides the tight correlation around the 1:1 correlation, we note two other populations in Figure \ref{fig:channel1density}---those lying below this line, having the detection statistic of the processed signal suppressed when compared to the original signal, and those lying above, conversely having the detection statistic augmented. The latter population will be discussed further in Section~\ref{sec:augmentation}.
 
\begin{figure*}[h!]
\centering
\includegraphics[width=0.7\textwidth]{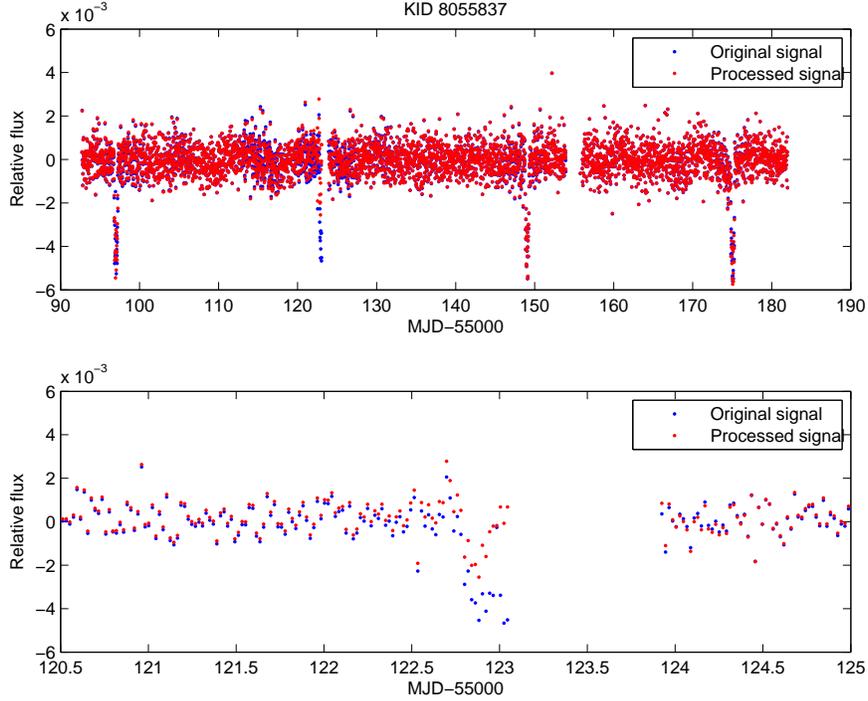}
\caption{\emph{Upper panel:} The Q3 flux time series for target KIC 8055837 on channel 1. The blue points are the detrended flux time series of the original flux time series from TPS, with the baseline signal for comparison injected at the end of the pipeline, and the red points are the same for the signal processed all the way through the pipeline. \emph{Lower panel:} The same as the upper panel, expanded to show the change in the injected signal; the edge correction before the long data gap has reduced the depth of the processed signal compared to the baseline signal.}
\label{fig:kid8055837}
\end{figure*}

There are 17 targets on this channel with significant suppression; we define `significant' suppression as having more than one cadence with a baseline SNR of $\ge10\sigma$ and a measured SNR of $<5\sigma$ for convenience in identifying outlier targets. For each channel, the number of targets with significant suppression is listed in Table \ref{tab:all80channels} as $N_{\rm supp}$. Of the 17 suppressed targets on channel 1, 11 have at least one transit injected near a data gap caused by a spacecraft re-pointing. Figure \ref{fig:kid8055837} shows an example, for target KID~8055837. There are four significant transit events in the flux time series shown in the upper panel; these are the injected transits. The second injected transit falls immediately before the first long data gap, which is due to the monthly data downlink from the spacecraft. As described in \citet{Tenenbaum2012a}, the data that fall immediately before and after a spacecraft re-pointing are fit and corrected with exponential terms to mitigate the impact of the thermal re-settling of the spacecraft. When a transit event occurs close to a re-pointing, the exponential fit will attempt to include the transit event in the correction, with the typical result that the transit signal strength is suppressed. On further examination of this and other channels, we find empirically that this occurs for transits that fall within 15 hours of the data gap caused by the spacecraft re-pointing. We also note that this suppression is almost exclusively found for transits with durations longer than $\sim$8~hours (i.e., transits with shorter durations do not affect enough of the cadences near the data gaps to influence the exponential fit). The suppression of transit signal strength near long data gaps is by far the dominant cause of suppression that we have noted, and is discussed further in Section~\ref{sec:suppression}. 

\begin{figure*}[h!]
\centering
\includegraphics[width=0.7\textwidth]{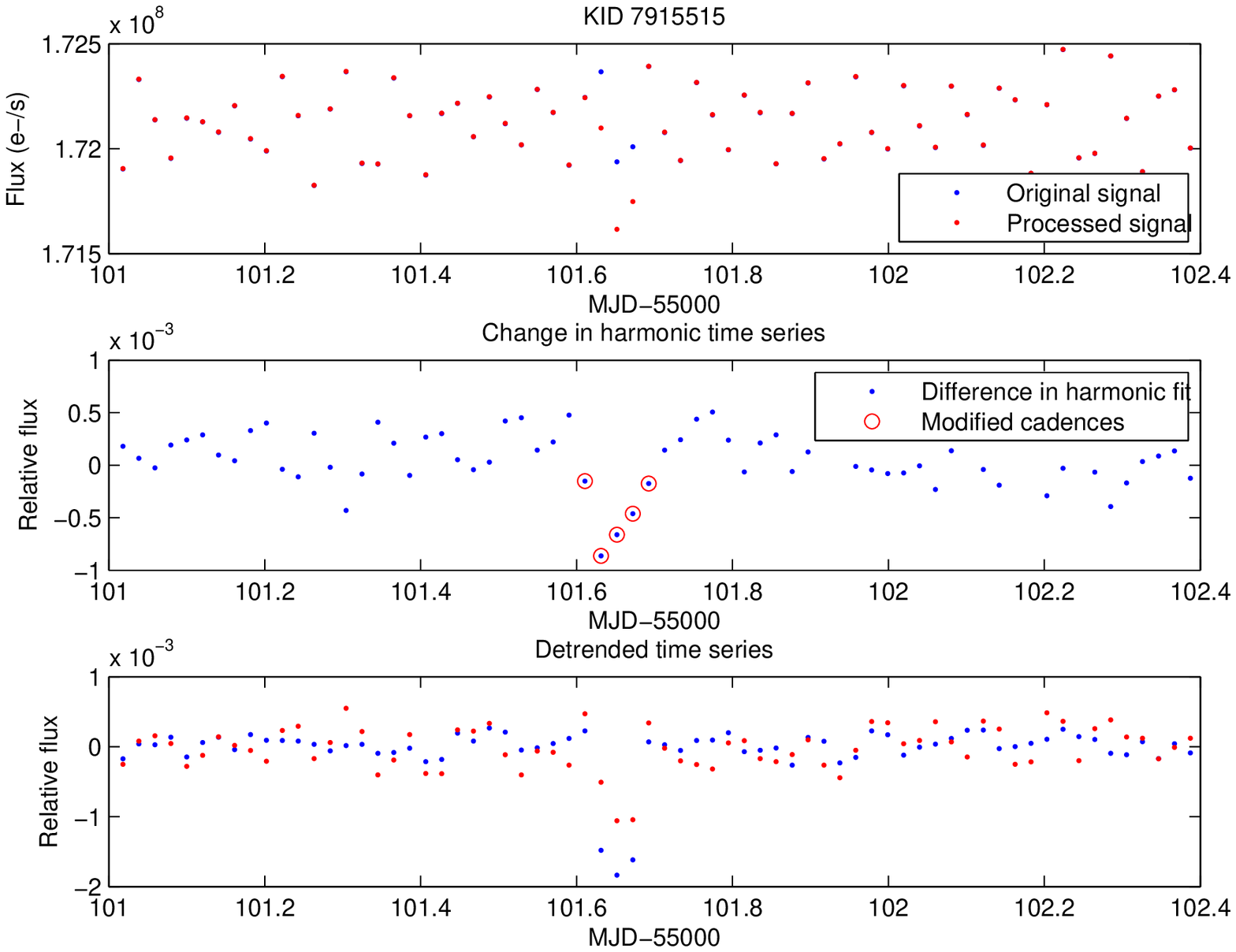}
\caption{\emph{Upper panel:} A small section of the Q3 flux time series for target KID 7915515 on channel 1. The blue points are the baseline PDC-corrected flux time series without the injected transits; the red points are the processed PDC-corrected flux time series with injected transits. The target exhibits high-frequency variability with an amplitude comparable to the depth of the injected transit, centred at MJD-55000=101.65. \emph{Middle panel:} The blue points show the difference in the harmonic fits to the flux time series in the upper panel. The points marked with red circles show the cadences modified by transit injection; these are the cadences with the largest differences in the harmonic fits. \emph{Lower panel:} The resulting detrended flux time series. The blue points show the baseline signal, and the red points show the processed signal. The processed transit depth is significantly shallower than the baseline depth.}
\label{fig:kid7915515}
\end{figure*}

\begin{figure*}[h!]
\centering
\includegraphics[width=0.7\textwidth]{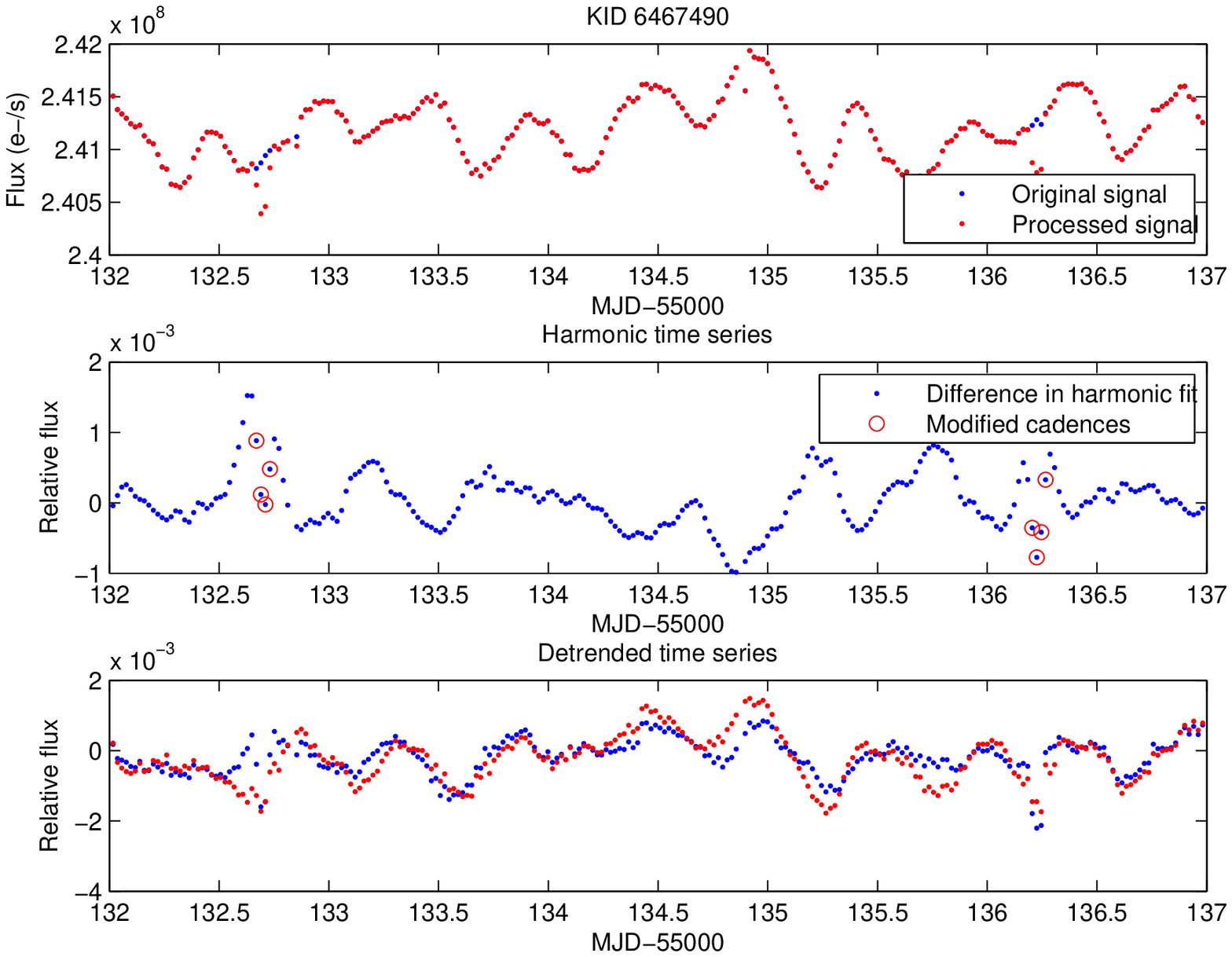}
\caption{The subsection of the Q3 flux time series for target KID 6777538 on channel 14, showing the affect of the harmonic fitter on the measured transit depth. See the caption of Figure \ref{fig:kid7915515} for details.}
\label{fig:kid6777538}
\end{figure*}

The transits in five of the 17 targets were suppressed by the harmonic fitter in TPS. Before the flux time series is whitened and searched for periodic signals, a sinusoidal harmonic filter is applied to remove periodic stellar activity, allowing the pipeline to search variable stars for transit signals \citep{Tenenbaum2012a}. In a small number of cases, the presence of transits affects the fitted harmonics, resulting in a different set of detection statistics. Figures \ref{fig:kid7915515} and \ref{fig:kid6777538} show examples of the harmonic fitter fitting a high frequency component to the transits themselves and attempting to filter them out, reducing their significance in the resulting flux time series; two of the five targets showed this behaviour by the harmonic fitter. We see this type of behaviour largely for variable stars where the flux time series already contains high-frequency components with a characteristic period $\sim$2--4 times the transit duration. For the remaining three targets, the harmonic fitter either fit no harmonics or a significantly reduced set of harmonics in the presence of the injected transits compared to the `clean' light curves, which had the effect of increasing the measured CDPP and therefore decreasing the significance of the detection statistics (which are normalised by CDPP). 

In the final case of significant signal suppression on channel 1, the outlier detection algorithm in PDC identified a different set of cadences as being `outliers' in the presence of injected transits, including several within 0.5 days of one of the injected transits. These outliers ultimately resulted in a measured SNR for this target which was suppressed compared to the baseline SNR. This was the only instance of the outlier detection algorithm resulting in a significantly suppressed signal, and therefore we do not believe this behaviour to be endemic in the pipeline.

\subsubsection{Channel 14}

\begin{figure*}[h!]
\centering
\includegraphics[width=0.45\textwidth]{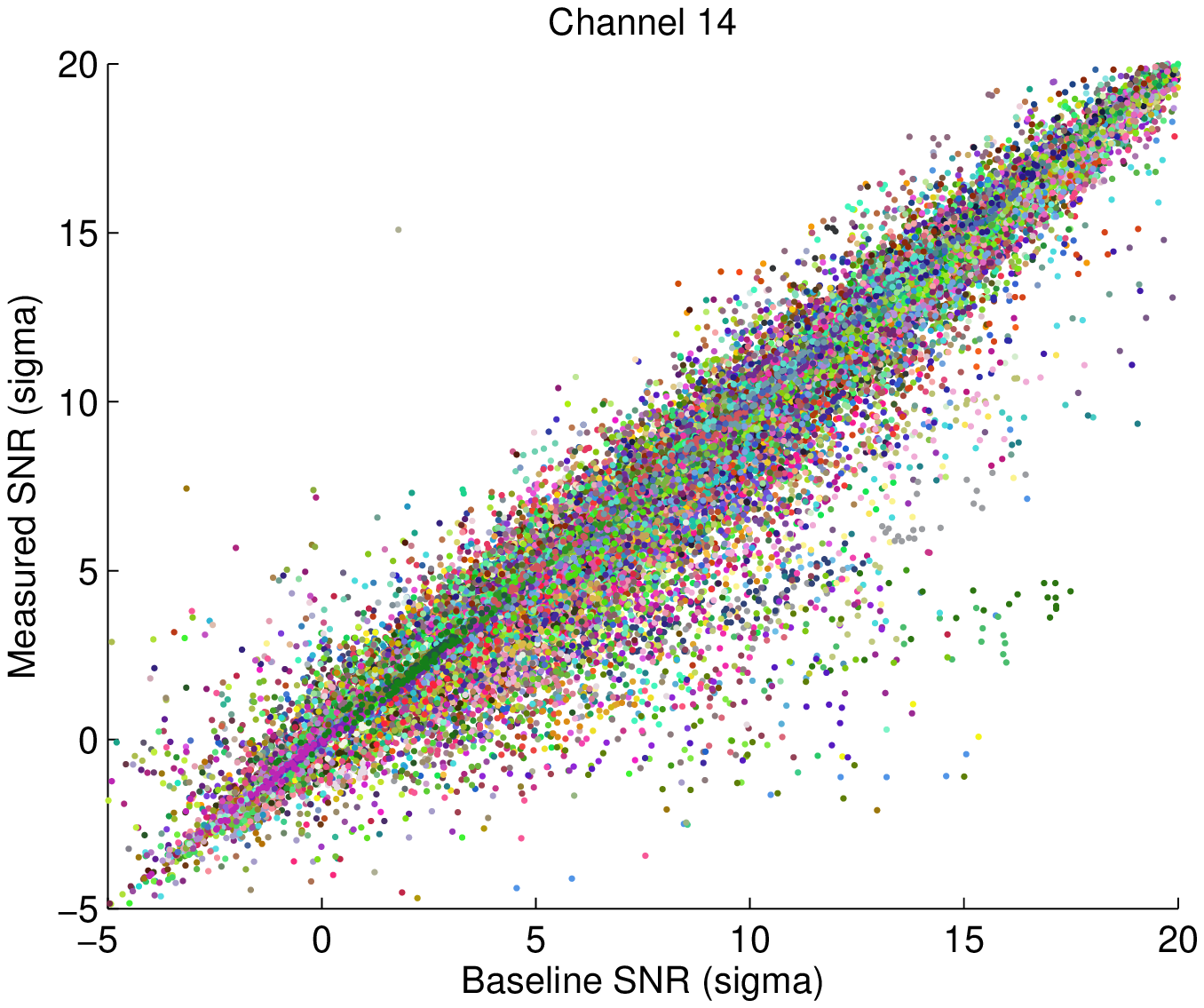}
\includegraphics[width=0.45\textwidth]{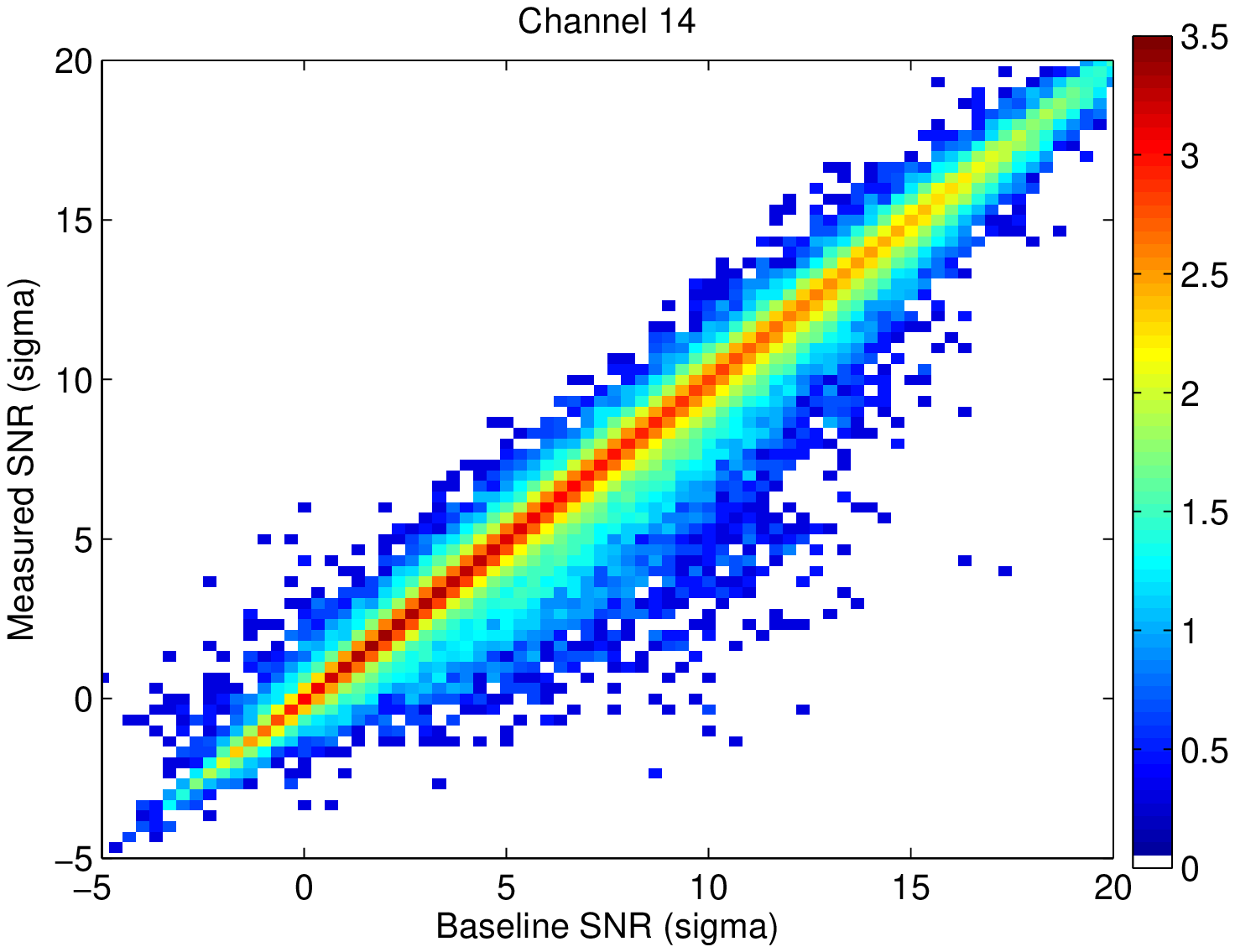}
\caption{The measured and baseline SNR for the 1510 targets on channel 14. See the Figure \ref{fig:channel1density} caption for details.}
\label{fig:channel14fits}
\end{figure*}

Figure \ref{fig:channel14fits} shows measured and baseline SNR for the 1510 targets on channel 14, in the same manner as Figure \ref{fig:channel1density}. The robust linear fit coefficients, as calculated for channel 1, were $a = 0.9965\pm0.0003$ and $b = -0.0126\pm0.0013$. We examine the distribution of the scaled residuals and find overall widths of 0.0287 for a Gaussian fit to the core ($|(MS-BS)/BS| < 0.06$) and 0.0294 for a Lorentzian fit to the core+wings ($|(MS-BS)/BS| < 0.4$), increasing from 0.0234 and 0.0233 respectively for baseline SNR values $>15\sigma$ to 0.0339 and 0.0402 for baseline SNR values $<0\sigma$. We again see the main population centred on the 1:1 line, with a second population of suppressed signals. We do not see a significant population of targets with augmented signals on this channel; overall, augmentation of signals appears much more rarely than suppression.

There are 25 targets on this channel with significant suppression, having more than one cadence  with a baseline SNR of $\ge10\sigma$ and a measured SNR of $<5\sigma$. Of these, 21 targets have transits injected adjacent to data gaps caused by spacecraft re-pointing, as discussed for channel 1. The remaining four targets were caused by the harmonic fitter in TPS being affected by the presence of the injected transits; Figure \ref{fig:kid6777538} shows an example from this channel.


\subsubsection{Channel 37}

\begin{figure*}[h!]
\centering
\includegraphics[width=0.45\textwidth]{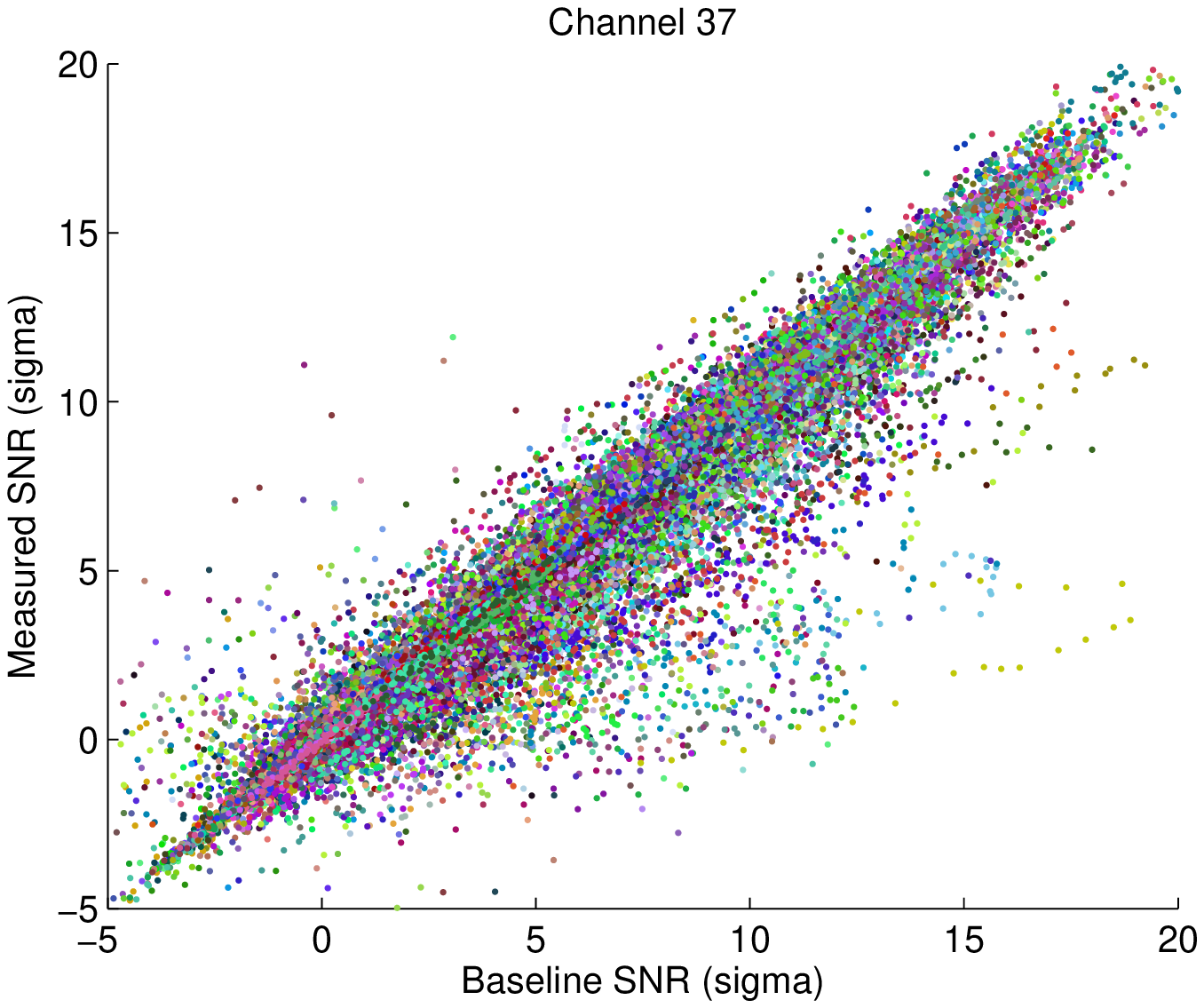}
\includegraphics[width=0.45\textwidth]{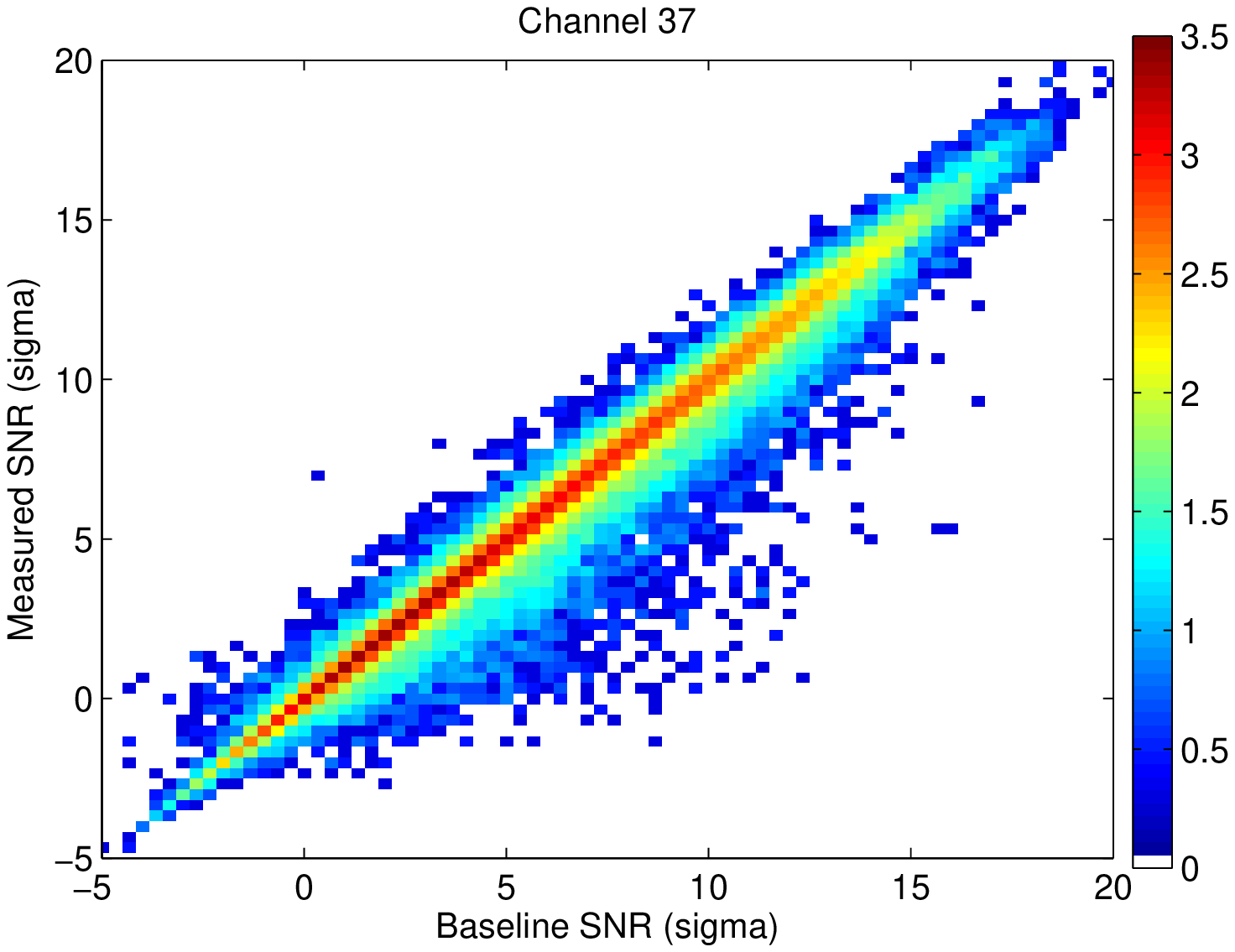}
\caption{The measured and baseline SNR for the 1448 targets on channel 37. See the Figure \ref{fig:channel1density} caption for details.}
\label{fig:channel37fits}
\end{figure*}

Figure \ref{fig:channel37fits} shows the measured and baseline SNR for the 1448 targets on channel 37, in the same manner as Figure \ref{fig:channel1density}. The robust linear fit coefficients, as calculated for channel 1, were $a = 0.9954\pm0.0003$ and $b = -0.0102\pm0.0012$. We examine the distribution of the scaled residuals and find overall widths of 0.0330 for a Gaussian fit to the core ($|(MS-BS)/BS| < 0.06$) and 0.0383 for a Lorentzian fit to the core+wings ($|(MS-BS)/BS| < 0.4$), increasing from 0.0300 and 0.0342 respectively for baseline SNR values $>15\sigma$ to 0.0354 and 0.0452 for baseline SNR values $<0\sigma$. The majority of the targets lie on the 1:1 line, with only a small number appearing to be systematically suppressed. We again do not see a significant population of targets with augmented signals on this channel.

There are 17 targets on this channel with significant suppression. Of these, 15 targets have transits injected adjacent to data gaps caused by spacecraft re-pointing, as discussed for channel 1. One of the remaining targets was caused by the harmonic fitter in TPS being altered by the presence of the injected transits. The final target was affected by the cosmic ray detection algorithm in the pipeline, as described in Section \ref{sec:distortion}, with two transits being `corrected' and therefore having their depth reduced. This was the only target found in this test to be affected in this way.

\subsubsection{Channel 64}
\label{sec:channel64}

\begin{figure*}[h!]
\centering
\includegraphics[width=0.45\textwidth]{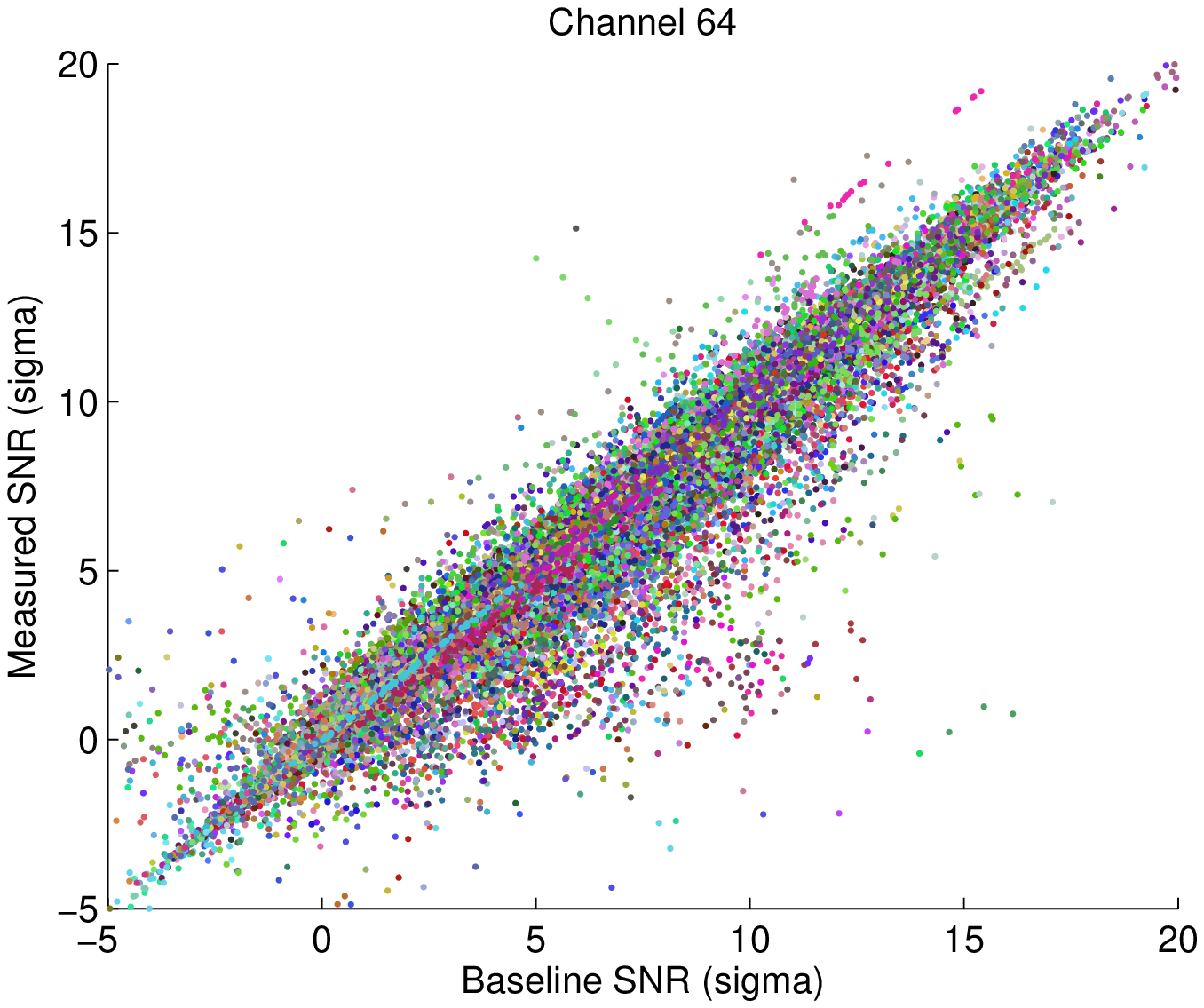}
\includegraphics[width=0.45\textwidth]{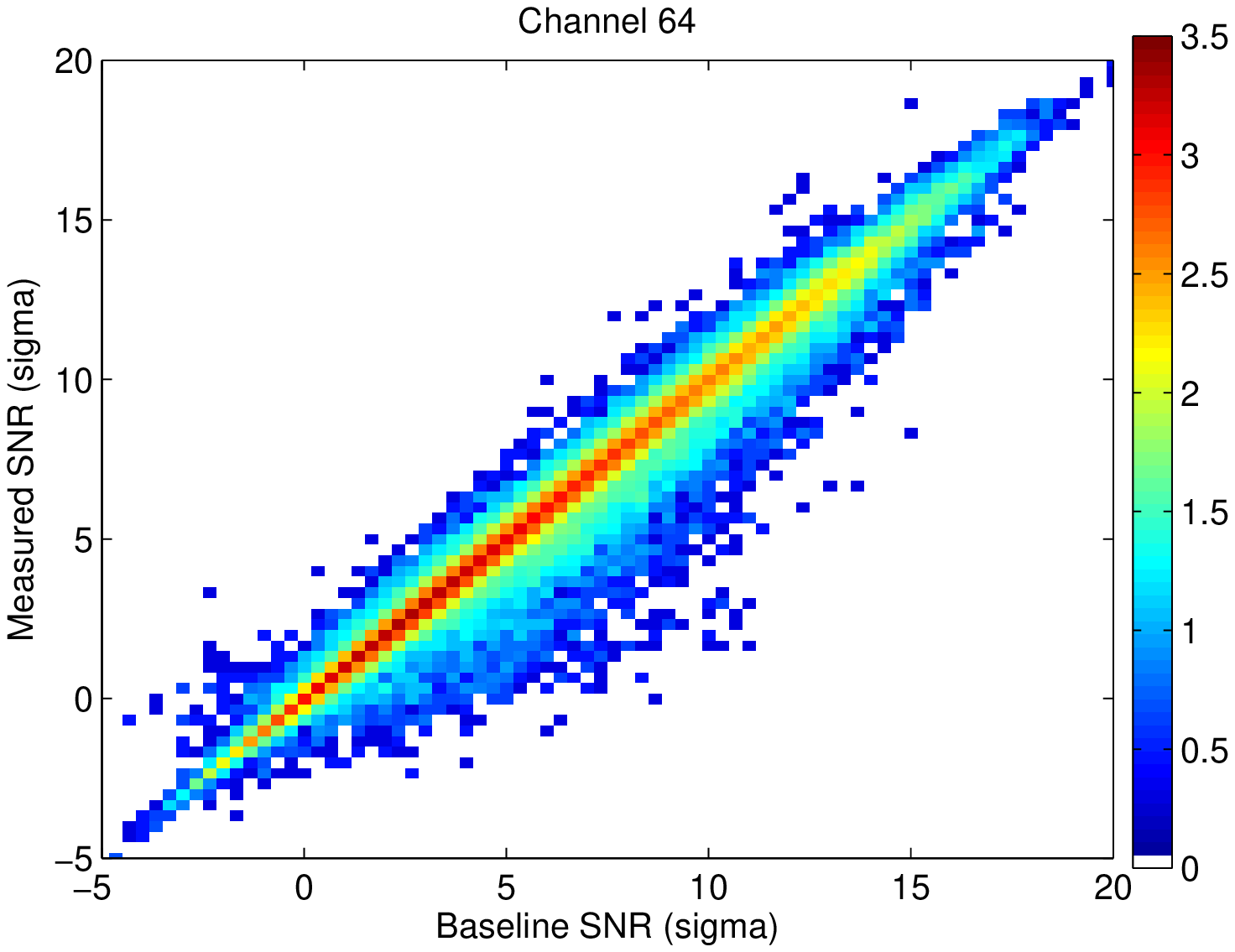}
\caption{The measured and baseline SNR for the 1148 targets on channel 64. See the Figure \ref{fig:channel1density} caption for details.}
\label{fig:channel64fits}
\end{figure*}

Finally, we examine the measured and baseline SNR for the 1148 targets on channel 64, shown in Figure \ref{fig:channel64fits}. The robust linear fit coefficients, as calculated for channel 1, were $a = 0.9970\pm0.0003$ and $b = -0.0151\pm0.0012$. We examine the distribution of the scaled residuals and find overall widths of 0.0306 for a Gaussian fit to the core ($|(MS-BS)/BS| < 0.06$) and 0.0333 for a Lorentzian fit to the core+wings ($|(MS-BS)/BS| < 0.4$), increasing from 0.0266 and 0.0275 respectively for baseline SNR values $>15\sigma$ to 0.0325 and 0.0355 for baseline SNR values $<0\sigma$. The majority of the targets lie on the 1:1 line, with only a small number appearing to be systematically suppressed. We again do not see a significant population of targets with augmented signals on this channel.

There are 7 targets with significant suppression, having more than one cadence  with an expected detection statistic of $\ge10\sigma$ and a measured detection statistic of $<5\sigma$. Of these, 5 targets have transits injected adjacent to long data gaps, as discussed for channel 1. One of the remaining targets was caused by the harmonic fitter in TPS being affected by the presence of the injected transits. 

The final target, KID~7510447, is a variable star showing spot modulation. The detection statistics were ultimately impacted by the PDC correction, as described in Section \ref{sec:augmentation}. Briefly, in this case the weight given to the priors in PDC for the correction to the flux time series with transits injected was much lower than the weight given to the priors for the `clean' flux time series. This leads to a larger amount of high-frequency noise being introduced to the flux time series with transits, increasing the measured CDPP and lowering the significance of the transit detection statistics. This was the only target for which this behaviour was observed; Section \ref{sec:augmentation} describes the handful of cases discovered showing the opposite behaviour.

\subsection{Overall signal recovery rate}

\begin{figure*}[h!]
\centering
\includegraphics[width=\textwidth]{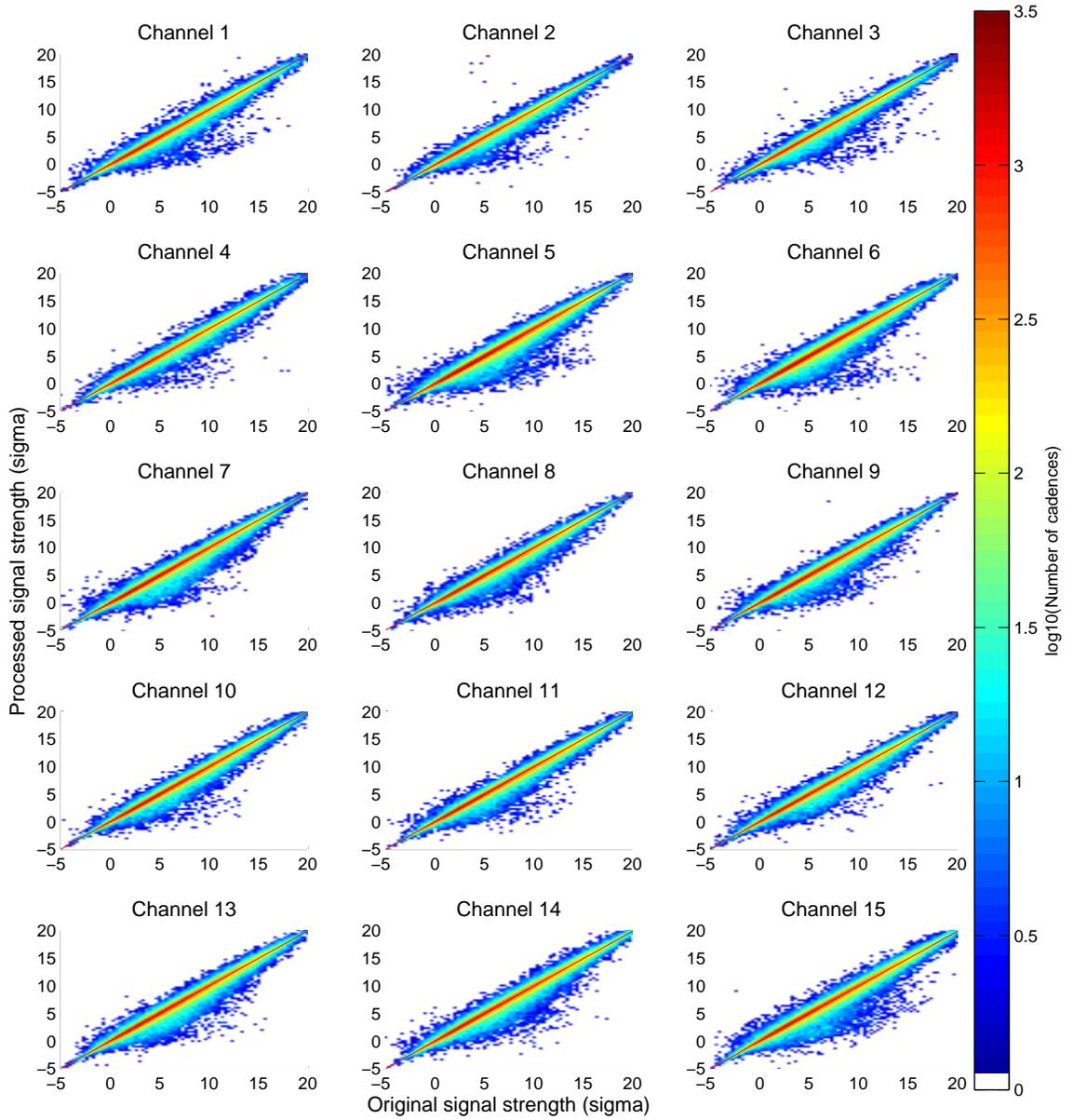}
\caption{First 15 channels.}
\label{fig:all80channels}
\end{figure*}

Figure \ref{fig:all80channels} shows the density plots of the measured SNR compared to the baseline SNR for the first 15 channels, illustrating that the overall response of each channel is largely the same; we examined the density plots for the remainder of the channels and see no departure from this pattern. Table \ref{tab:all80channels} summarises the fits for the 80 channels, showing the coefficients of the robust linear fit and the numbers of targets analysed in each case. If the signal were being perfectly preserved by the pipeline, we would expect a 1:1 correlation between the measured and baseline SNR; in fact we are extremely close to this for every channel. The median recovery rate across all channels is:

\begin{equation}
MS = 0.9973(\pm0.0011)\times BS - 0.0151(\pm0.0049)
\end{equation}

This corresponds to an average measured SNR of 9.96$\sigma$ for a 10$\sigma$ signal, a $\sim$0.4\% reduction. Across the 80 channels the linear term ranges from 0.9936--1.0004 and the constant term from -0.0266--0.0012. The width of this distribution is also shown for each channel in Table \ref{tab:all80channels}, for the Gaussian and Lorentzian fits to the core and core+wings respectively. The median Gaussian HWHM across all channels is 0.0264, with individual channels ranging from 0.0246--0.0360. The median Lorentzian HWHM across all channels is 0.0256, with individual channels ranging from 0.0220--0.0461. Therefore across the full focal plane, simulated transit signals are typically recovered to within 2--4\% of their expected SNR.

\clearpage

\begin{center}
\begin{longtable}[h!]{|cccccccc|}
\caption{SNR recovery for each channel. \# Stars is the final number of targets used in the analysis. $a$ and $b$ are coefficients of the fit $MS = a \times BS + b$, with 1$\sigma$ errors shown for the last two significant digits. HWHM$_{\rm Gauss}$ and HWHM$_{\rm Lor}$ are the half-width half-maxima of Gaussian and Lorentzian fits to the residuals, as described in the text. N$_{\rm supp}$ and N$_{\rm aug}$ are the numbers of targets showing significant suppression or augmentation of their detection statistics respectively, as defined in Sections \ref{sec:channel1} and \ref{sec:augmentation}.}
\label{tab:all80channels} \\
\hline
\small{Channel} & \small{\# Stars} & \small{$a$ ($\pm \Delta a$)} & \small{$b$  ($\pm \Delta b$)} & \small{HWHM$_{\rm Gauss}$} & \small{HWHM$_{\rm Lor}$} & \small{N$_{\rm supp}$} & \small{N$_{\rm aug}$} \\
\hline
\endfirsthead
\multicolumn{8}{c}{{\small \tablename\ \thetable{} -- continued from previous page}} \\
\hline 
\small{Channel} & \small{\# Stars} & \small{$a$ ($\pm \Delta a$)} & \small{$b$  ($\pm \Delta b$)} & \small{HWHM$_{\rm Gauss}$} & \small{HWHM$_{\rm Lor}$} & \small{N$_{\rm supp}$} & \small{N$_{\rm aug}$} \\
\hline
\endhead
\hline \multicolumn{8}{c}{{Continued on next page}} \\ \hline
\endfoot
\hline \hline
\endlastfoot
1 & 1130 & -0.0218(15) & 0.9970(03) & 0.0315 & 0.0356 & 13 & 0 \\ 
2 & 768 & -0.0175(16) & 0.9936(03) & 0.0322 & 0.0372 & 9 & 1 \\ 
3 & 815 & -0.0156(20) & 0.9965(03) & 0.0315 & 0.0358 & 10 & 1 \\ 
4 & 946 & -0.0266(16) & 0.9962(03) & 0.0313 & 0.0352 & 9 & 0 \\ 
5 & 2316 & -0.0164(06) & 0.9985(03) & 0.0254 & 0.0235 & 22 & 0 \\ 
6 & 2192 & -0.0177(08) & 0.9977(03) & 0.0253 & 0.0239 & 17 & 0 \\ 
7 & 1660 & -0.0184(09) & 0.9977(03) & 0.0267 & 0.0261 & 9 & 0 \\ 
8 & 1242 & -0.0169(11) & 0.9988(03) & 0.0255 & 0.0241 & 8 & 0 \\ 
9 & 1568 & -0.0101(08) & 0.9975(03) & 0.0252 & 0.0232 & 13 & 1 \\ 
10 & 1506 & -0.0119(10) & 0.9985(03) & 0.0246 & 0.0220 & 13 & 0 \\ 
11 & 1391 & -0.0185(10) & 0.9989(03) & 0.0258 & 0.0241 & 12 & 0 \\ 
12 & 1248 & -0.0151(11) & 0.9973(03) & 0.0261 & 0.0251 & 10 & 0 \\ 
13 & 1605 & -0.0227(12) & 0.9984(03) & 0.0286 & 0.0296 & 20 & 0 \\ 
14 & 1510 & -0.0126(13) & 0.9965(03) & 0.0287 & 0.0294 & 21 & 0 \\ 
15 & 1628 & -0.0096(12) & 0.9960(03) & 0.0297 & 0.0314 & 25 & 1 \\ 
16 & 1534 & -0.0185(14) & 0.9972(03) & 0.0306 & 0.0335 & 21 & 0 \\ 
17 & 1480 & -0.0140(08) & 0.9991(03) & 0.0265 & 0.0256 & 9 & 0 \\ 
18 & 1860 & -0.0144(08) & 0.9974(03) & 0.0252 & 0.0234 & 6 & 0 \\ 
19 & 1508 & -0.0187(09) & 0.9982(03) & 0.0262 & 0.0245 & 13 & 1 \\ 
20 & 1386 & -0.0133(10) & 0.9973(03) & 0.0271 & 0.0264 & 12 & 0 \\ 
21 & 1520 & -0.0155(10) & 0.9977(03) & 0.0273 & 0.0269 & 10 & 0 \\ 
22 & 1687 & -0.0139(08) & 0.9973(03) & 0.0255 & 0.0241 & 7 & 0 \\ 
24 & 1649 & -0.0192(09) & 0.9983(03) & 0.0261 & 0.0244 & 5 & 0 \\ 
25 & 1431 & -0.0166(09) & 0.9982(03) & 0.0251 & 0.0228 & 7 & 0 \\ 
26 & 1134 & -0.0154(09) & 0.9995(03) & 0.0253 & 0.0228 & 5 & 0 \\ 
27 & 1300 & -0.0162(10) & 0.9971(03) & 0.0260 & 0.0243 & 6 & 0 \\ 
28 & 1388 & -0.0109(10) & 0.9972(03) & 0.0254 & 0.0233 & 12 & 1 \\ 
29 & 1081 & -0.0108(14) & 0.9964(03) & 0.0296 & 0.0312 & 10 & 0 \\ 
30 & 957 & -0.0099(13) & 0.9952(03) & 0.0289 & 0.0301 & 7 & 1 \\ 
31 & 906 & -0.0158(14) & 0.9979(03) & 0.0276 & 0.0280 & 5 & 0 \\ 
32 & 852 & -0.0063(13) & 0.9970(03) & 0.0259 & 0.0246 & 7 & 0 \\ 
33 & 1853 & -0.0128(09) & 0.9974(03) & 0.0270 & 0.0265 & 24 & 0 \\ 
34 & 2200 & -0.0149(08) & 0.9983(03) & 0.0259 & 0.0246 & 22 & 0 \\ 
35 & 1893 & -0.0159(11) & 0.9964(03) & 0.0302 & 0.0327 & 21 & 0 \\ 
36 & 1600 & -0.0154(12) & 0.9985(03) & 0.0316 & 0.0354 & 14 & 0 \\ 
37 & 1683 & -0.0102(12) & 0.9954(03) & 0.0330 & 0.0383 & 16 & 0 \\ 
38 & 1741 & -0.0095(11) & 0.9950(03) & 0.0325 & 0.0373 & 13 & 0 \\ 
39 & 1765 & -0.0098(09) & 0.9970(03) & 0.0298 & 0.0317 & 19 & 0 \\ 
40 & 1741 & -0.0177(11) & 0.9957(03) & 0.0327 & 0.0381 & 9 & 0 \\ 
41 & 1413 & -0.0126(12) & 0.9954(03) & 0.0313 & 0.0346 & 14 & 0 \\ 
42 & 1313 & -0.0140(11) & 0.9965(03) & 0.0299 & 0.0322 & 12 & 0 \\ 
43 & 1485 & -0.0089(12) & 0.9956(03) & 0.0302 & 0.0327 & 12 & 0 \\ 
44 & 1448 & 0.0012(12) & 0.9942(03) & 0.0351 & 0.0429 & 4 & 0 \\ 
46 & 1249 & -0.0178(11) & 0.9986(03) & 0.0273 & 0.0269 & 11 & 0 \\ 
47 & 1195 & -0.0173(10) & 0.9978(03) & 0.0254 & 0.0241 & 11 & 0 \\ 
48 & 1153 & -0.0229(11) & 0.9983(03) & 0.0248 & 0.0225 & 9 & 0 \\ 
49 & 1134 & -0.0167(11) & 0.9990(03) & 0.0253 & 0.0238 & 8 & 0 \\ 
50 & 1082 & -0.0174(11) & 0.9973(03) & 0.0257 & 0.0248 & 7 & 0 \\ 
51 & 850 & -0.0082(16) & 0.9964(03) & 0.0256 & 0.0243 & 14 & 0 \\ 
52 & 887 & -0.0104(14) & 0.9982(03) & 0.0248 & 0.0228 & 12 & 0 \\ 
53 & 1445 & -0.0130(15) & 0.9971(03) & 0.0352 & 0.0439 & 21 & 0 \\ 
54 & 1391 & -0.0134(16) & 0.9965(03) & 0.0360 & 0.0461 & 20 & 2 \\ 
55 & 1462 & -0.0105(15) & 0.9952(03) & 0.0354 & 0.0451 & 22 & 1 \\ 
56 & 933 & -0.0034(19) & 0.9952(03) & 0.0337 & 0.0402 & 12 & 0 \\ 
57 & 1343 & -0.0139(09) & 0.9982(03) & 0.0259 & 0.0246 & 7 & 1 \\ 
58 & 1417 & -0.0041(08) & 0.9981(03) & 0.0257 & 0.0236 & 6 & 0 \\ 
59 & 1765 & -0.0199(08) & 0.9980(03) & 0.0260 & 0.0249 & 7 & 1 \\ 
60 & 1303 & -0.0148(10) & 0.9979(03) & 0.0277 & 0.0275 & 10 & 0 \\ 
61 & 1225 & -0.0092(13) & 0.9964(03) & 0.0307 & 0.0335 & 12 & 0 \\ 
63 & 1160 & -0.0129(12) & 0.9962(03) & 0.0296 & 0.0314 & 12 & 0 \\ 
64 & 1148 & -0.0151(12) & 0.9970(03) & 0.0306 & 0.0333 & 8 & 0 \\ 
65 & 1023 & -0.0238(11) & 1.0004(03) & 0.0256 & 0.0235 & 15 & 0 \\ 
66 & 1309 & -0.0125(09) & 0.9977(03) & 0.0263 & 0.0250 & 6 & 0 \\ 
68 & 963 & -0.0173(09) & 0.9990(03) & 0.0249 & 0.0232 & 9 & 0 \\ 
69 & 885 & -0.0140(13) & 0.9973(03) & 0.0255 & 0.0240 & 9 & 0 \\ 
70 & 746 & -0.0166(13) & 0.9982(03) & 0.0263 & 0.0249 & 6 & 0 \\ 
71 & 663 & -0.0085(16) & 0.9959(03) & 0.0256 & 0.0246 & 5 & 0 \\ 
72 & 635 & -0.0178(17) & 0.9980(03) & 0.0260 & 0.0247 & 7 & 0 \\ 
73 & 900 & -0.0133(16) & 0.9960(03) & 0.0291 & 0.0307 & 11 & 0 \\ 
74 & 1046 & -0.0101(14) & 0.9968(03) & 0.0313 & 0.0347 & 15 & 0 \\ 
75 & 1073 & -0.0072(13) & 0.9972(03) & 0.0289 & 0.0302 & 10 & 0 \\ 
76 & 1018 & -0.0123(14) & 0.9964(03) & 0.0299 & 0.0325 & 11 & 0 \\ 
77 & 1173 & -0.0190(12) & 0.9980(03) & 0.0255 & 0.0242 & 14 & 0 \\ 
78 & 1132 & -0.0222(12) & 0.9985(03) & 0.0261 & 0.0254 & 9 & 0 \\ 
79 & 903 & -0.0164(13) & 0.9983(03) & 0.0269 & 0.0265 & 11 & 0 \\ 
80 & 893 & -0.0160(15) & 0.9964(03) & 0.0261 & 0.0256 & 3 & 0 \\ 
81 & 749 & -0.0157(27) & 0.9974(04) & 0.0250 & 0.0229 & 6 & 0 \\ 
82 & 885 & -0.0230(13) & 0.9987(03) & 0.0247 & 0.0222 & 8 & 0 \\ 
83 & 684 & -0.0212(13) & 0.9979(03) & 0.0249 & 0.0237 & 0 & 0 \\ 
84 & 622 & -0.0200(17) & 0.9969(03) & 0.0254 & 0.0242 & 6 & 1 \\ 
\hline
\end{longtable}
\end{center}

\section{Discussion}
\label{sec:discussion}

\subsection{Signal suppression}
\label{sec:suppression}

We examined a set of targets where the measured signal strength was suppressed compared to the expected value. Any systematic reduction in signal strength by the processing pipeline would have significant consequences when determining the population of planets detectable in the \kepler\ data. We find that $\sim$0.9\% of targets experience significant suppression of at least one transit, with more than one modified cadence having a measured SNR of $<5\sigma$ with an expected, baseline SNR of $>10\sigma$.

In the large majority of cases ($\sim$80\%) the cause of the suppression was the target having a long-duration ($>8$ hours) transit injected near to (within 15 hours of) a long data gap. For these targets, the remaining injected transits were recovered as normal (i.e., even in the targets showing significant suppression, the majority of the transits are not suppressed). The current algorithm in TPS which corrects for thermal systematics after spacecraft re-pointing is impacted by the presence of transits in the flux time series window used in the correction. In Q3, there were six `edges' to long data gaps, at the start and end of each month. It is straight-forward to re-calculate the correlations described in Section \ref{sec:results} after removing cadences that fall within 15 hours of these edges. Overall we find an improved recovery rate of $MS = 0.9972(\pm0.0012)\times BS - 0.0036(\pm0.0011)$; the slope is essentially identical, and the offset is significantly reduced, indicating that a large fraction of the original offset is due to these suppressed transits. On average the width of the distribution of the scaled residuals did not change when removing cadences near the gaps: the median Gaussian and Lorentzian HWHM across all channels changed from 0.0264 and 0.0256 respectively to 0.0263 and 0.0254 respectively.

In quarters with additional spacecraft re-pointings, due to safe modes, for instance, there may be additional `edges'. We encourage readers to account for these windows of reduced signal recovery when determining the detectability of a given planet candidate, by using a reduced observation baseline when calculating phase coverage. Transits of Earth-size planets around sun-like stars average $\sim$10~hours, and thus could experience this effect if they fall close to a long data gap. The \kepler\ pipeline applies a set of de-emphasis weights to points falling within two days of a data gap when calculating the significance of signals in TPS. Most spacecraft events that do not involve re-pointing typically produce short gaps (such as losses of fine point, reaction wheel momentum desaturations, etc) that do not impact signal strength recovery.

In a further $\sim$10\% of cases, injected transits had their recoverability impacted by the harmonic fitter in TPS, either by being directly fit and removed by the fitter, or by the change in fitted harmonics  increasing the measured CDPP relative to the `clean' light curves. This effect was largely seen for stars that already exhibited high-frequency variability (with periods of hours to tens of hours). An important note here is that the injected transit epochs have an artificial separation of 50 times the transit duration, representing a `duty cycle', or fraction of the phased light curve occupied by the transit, of 0.02. We expect and have observed the behaviour of the harmonic fitter to be highly dependent on the duty cycle of the transit signal; the higher the duty cycle, the more likely the harmonic fitter will be able to fit the transit with a Fourier series of sine waves. Although a duty cycle of 0.02 does represent real physical parameter space for planets in short-period orbits, at longer and more interesting periods the duty cycle falls by an order of magnitude, and we expect the harmonic fitter to have a significantly smaller impact. We will examine this behaviour in a subsequent analysis, where we inject transits with realistic separations into longer observation baselines.

\subsection{Signal augmentation}
\label{sec:augmentation}

\begin{figure*}[h!]
\centering
\includegraphics[width=\textwidth]{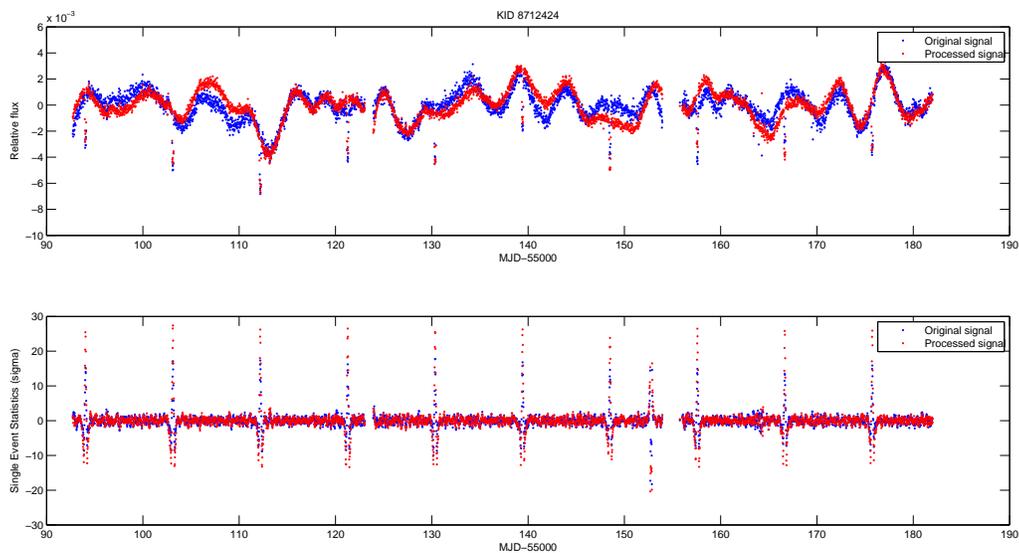}
\caption{\emph{Upper panel:} The Q3 flux time series for target KID 8712424 on channel 1. The blue points are the detrended flux time series from TPS of the original flux time series, with the simulated transits injected at the end of the pipeline, and the red points are the same for the flux time series with the simulated transits injected into the calibrated pixels. \emph{Lower panel:} The single event detection statistics for the two flux time series in the upper panel. The measured SNR of the injected transits (in red) is higher than the baseline SNR (in blue).}
\label{fig:kid8712424}
\end{figure*}

A very small number of targets experienced systematic augmentation of their detection statistics during processing; there are only 13 targets across the full 80 channels with more than one modified cadence having a baseline SNR of $\le5\sigma$ and a measured SNR of $>10\sigma$ ($N_{\rm aug}$ in Table \ref{tab:all80channels}). We find that these are almost all variable stars showing evidence of spot modulation. Figure~\ref{fig:kid8712424} shows an example from channel 2. In these cases, PDC incorrectly gave the priors a very low weight when processing the clean light curves, introducing high-frequency noise in the course of reducing the overall root-mean-square deviation, as described by \citet{Stumpe2012}. This leads to an increase in the measured CDPP for the original flux time series, and a correspondingly lower detection statistic, shown in the lower panel of Figure~\ref{fig:kid8712424}. This behaviour was corrected in the final version of SOC 8.3 and we do not expect future augmentation of signals as a result.


\subsection{Correlations of signal recovery with injected transit properties}

As noted in the previous discussion, there are some regions of parameter space where pipeline processing can behave significantly differently, such as long-duration transits falling near long data gaps, or transits in targets with high-frequency variability. Here we examine the overall dependence of the signal recoverability as a function of the injected transit attributes, in particular transit duration and transit SNR.

\begin{figure*}
\centering
\includegraphics[width=0.7\textwidth]{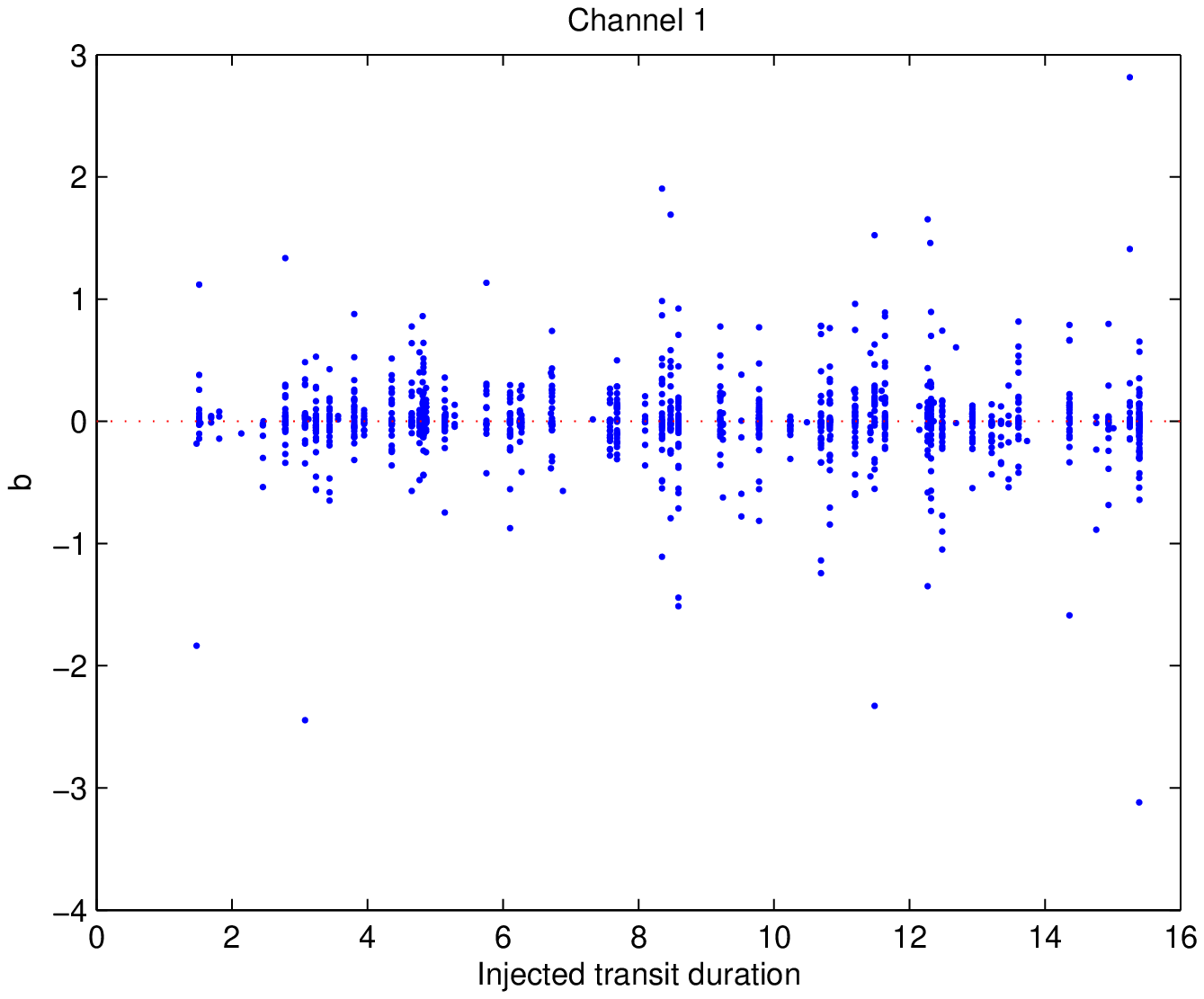}
\caption{The distribution of the constant coefficient, $b$, in the robust fit with injected transit duration.}
\label{fig:channel1dependence_duration}
\end{figure*}

For a given channel, we find the robust linear correlation between the measured detection statistics and the baseline detection statistics for each target \emph{individually}, again of the form $MS = a \times BS + b$. We then measure the robust linear correlation between the coefficients of those fits, $a$ and $b$, and the injected transit duration and injected transit SNR of the target. Figure \ref{fig:channel1dependence_duration} shows an example for channel 1, comparing the constant coefficient, $b$, found for each target to the injected transit duration for that target. There is no obvious correlation in the figure, and indeed we measure an insignificant correlation of -0.0022$\pm$0.0012. We measure the correlation of $b$ with duration for each channel, and find the correlations range from -0.0057 to 0.0003, with a median correlation of -0.0028, a standard deviation of 0.0011 across all channels and an average error per channel of 0.0014. For the linear coefficient, $a$, we find even less significance in the correlation, with correlations across the channels ranging from -0.00058 to 0.00067, a median correlation of 0.00010, a standard deviation of 0.00020 across all channels and an average error per channel of 0.00080. Therefore we find no significant correlation between the fidelity of a processed signal and its duration. The long-duration transits injected near long data gaps are clearly rare enough as to not influence the average detection statistics, which is useful to note for planet completeness requirements.

We perform the same analysis for the injected transit SNR, and again find no significant ($>3\sigma$) correlation between the fidelity of the measured SNR and the baseline SNR (i.e., low-significance transits are no more likely to be impacted by pipeline processing than high-significance transits, between 2$\sigma$ and 20$\sigma$). Formally, the median correlation of $b$ across all channels is 0.00230, with a standard deviation of 0.00089 and an average error per channel of 0.00087. The median correlation of $a$ across all channels is -0.00028, with a standard deviation of 0.00016 and an average error per channel of 0.00060. This is also qualitatively evident from Figure \ref{fig:channel1resids}, where the scaled residuals are shown for different bins of baseline SNR---there is no shift in the centre of the distribution with decreasing baseline SNR. There is, however, a widening of the distribution with lower baseline SNR, indicating that although there is no systematic change in the average measured SNR, the probability of measuring a SNR significantly different from the baseline SNR increases at lower baseline SNR.

\section{Conclusions and further work}
\label{sec:conclusion}

For a given transit signal, the \kepler\ pipeline has extremely high fidelity in reproducing the expected \emph{single} event detection statistics. The average recovery of an individual transit event is $MS = 0.9973(\pm0.0012)\times BS - 0.0151(\pm0.0049)$, where $MS$ is the measured SNR and $BS$ is the baseline, or expected, SNR. We have characterised the pipeline for single transit events from the summation of the pixels into a flux time series, through the systematic error correction, to the final detrending and whitening before the periodic signal search. 

This initial analysis was limited in scope to a single quarter of observation. This necessitated injecting the simulated transit signals with smaller separations in transit epoch than would be physically observed, in order to maximise the number of statistical tests that could be performed per target. Although we attempted to mitigate the influence of this change on the results, several effects that we have discussed (i.e., the SPSD detector, the behaviour of the harmonic fitter, and the influence of transits on the weight assigned to the priors in PDC) may not be reproducing the real performance of the pipeline. In the latter two cases, we expect the incidence rate of transit suppression to be lower with wider separations, in which case what we report here should be considered the worst-case scenario.

We conclude therefore that the pipeline does not systematically perturb the signal strength of individual transit events. This has positive implications for the planet sample completeness---for teams performing independent searches for periodic transit signals in the flux time series produced by the pipeline, this initial work provides sufficient information to move forward with determining the completeness of their own generated planet candidate set. For analysis of the \kepler\ planet candidate set, this work provides reassurance that no systematic reduction in transit signal strength is introduced by the pipeline, and that any significant incompleteness produced by the pipeline would therefore be localised to the periodic transit search.

In order to confirm our expectations, and more importantly to examine the performance of the pipeline through the entire analysis, including the periodic transit search (folding of the single event detection statistics (measured SNR) to create multiple event statistics), and the validation of the signal detection in DV, we need to perform an analysis similar to the one described in this work, but with realistic, physical separations of the transit epochs in time. To accommodate this wider separation in time, our next analysis will be performed on a significantly longer observation baseline.

We also plan to examine the performance of DV at recognising false positive signals. The ability to inject the simulated transit events at the pixel level allows us to introduce spatial offsets in the source of the transit signal. In this fashion we can mimic background eclipsing binary stars, or small biases in the measurements of the position of the targets.

\acknowledgments

Funding for the \kepler\ Discovery Mission is provided by NASA's Science Mission Directorate. 






\clearpage
\end{document}